\documentclass[]{iopart}

\usepackage{amsfonts}
\usepackage{amstext}
\usepackage{amssymb}
\usepackage{bm}
\usepackage{cite}
\usepackage{dcolumn}
\usepackage{graphicx}
\usepackage{graphics}
\usepackage[utf8]{inputenc}
\usepackage{latexsym}
\usepackage{rotating}
\usepackage{url}
\usepackage[colorlinks=true]{hyperref}
\usepackage{xspace} 
\usepackage[usenames]{color}
\usepackage{booktabs}

\expandafter\let\csname equation*\endcsname\relax
\expandafter\let\csname endequation*\endcsname\relax
\usepackage{amsmath}

\begin{document}

\title[Testing Gravity with Black Hole Shadow Subrings]{Testing Gravity with Black Hole Shadow Subrings}

\author{%
Dimitry~Ayzenberg
}

\address{Theoretical Astrophysics, Eberhard-Karls Universit{\"a}t T{\"u}bingen, D-72076 T{\"u}bingen, Germany}

\date{\today}

\begin{abstract} 

The black hole shadow, first observed by the Event Horizon Telescope in 2017, is the newest method for studying black holes and understanding gravity. Much work has gone into understanding the shadow of a Kerr black hole, including all of the complex astrophysics of the accretion disk, and there are numerous studies of the ideal shadow in non-Kerr black holes and exotic compact objects. This paper presents one of the first studies of the black hole shadow of non-Kerr black holes when the illumination source is an accretion disk. In particular, the ability of current and future very long baseline interformeters to estimate the physical parameters of the black hole spacetime and accretion disk is investigated using two different parametrized black hole metrics that encode a number of possible deviations from Kerr. Both the full shadow image and the individual subrings of the shadow are analyzed as the higher order subrings are weakly dependent on the disk physics and may be a more viable observable for studying the spacetime. The results suggest that with current telescope capabilities and any future earth-based telescopes it will be quite difficult to place strong constraints on departures from the Kerr spacetime, primarily due to the low resolution and strong degeneracies between the spacetime parameters. More optimistically, space-based interferometers may be capable of testing the Kerr nature of black holes and general relativity to comparable or better precision than is currently possible with other observations.

\end{abstract}

\submitto{\CQG}

\maketitle

\section{Introduction}

The 2017 observation by the Event Horizon Telescope (EHT) of the central supermassive black hole in the galaxy M87~\cite{2019ApJ...875L...1E} was the first of its kind. The EHT observed what is known as the black hole shadow~\cite{2019ApJ...875L...6E}, an observable unique to black holes (and possibly other exotic compact objects) that is due to the existence of an event horizon and/or a photon sphere. The event horizon is a surface of no return and is unique to black holes, while the photon sphere is the surface of unstable spherical photon orbits below which orbits are not possible and is not necessarily unique to black holes. That the black hole shadow is a consequence of the event horizon and photon sphere makes it a strong-field gravity observable,~i.e.~very near to the event horizon where observational tests have only begun in the past decade or so. Such strong-field observables make it possible to study the spacetime of black holes and gain a better understanding of gravity.

One of the primary goals of any black hole observation is to test the {\it Kerr hypothesis},~i.e.~that the spacetime of all astrophysical (uncharged), isolated, stationary, and axisymmetric black holes is given by the Kerr solution. The Kerr solution is fully determined by two parameters, the black hole mass $M$ and the black hole spin angular momentum $\vec J$. Under general relativity and some modified gravity theories the Kerr hypothesis holds~\cite{Psaltis:2007cw}, while in many others it does not~\cite{Yunes:2013dva}. Even within general relativity, it is possible black holes are not as isolated as we assume and are surrounded and influenced by other fields~\cite{PhysRevLett.112.221101}, in which case black holes are certainly not described by the Kerr metric. Additionally, the objects we assume to be black holes could instead be some other exotic compact object~\cite{Cardoso:2019rvt}. Observations of the black hole shadow can help to test all these various proposals for what we call black holes and determine the correct description for gravity.

\begin{figure*}[hpt]
\includegraphics[width=0.5\columnwidth{},clip=true]{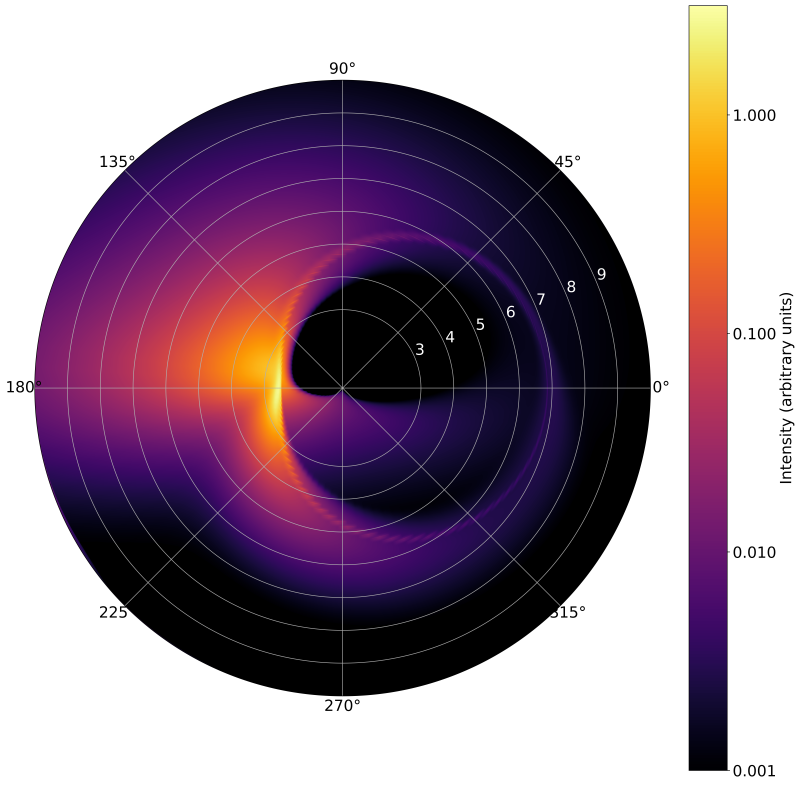}
\includegraphics[width=0.5\columnwidth{},clip=true]{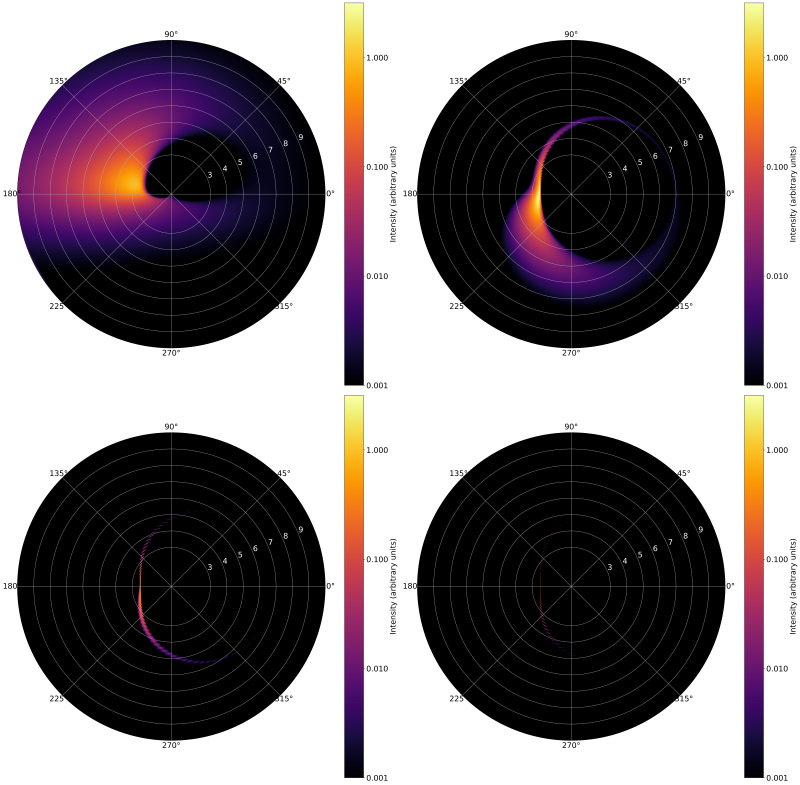}
\caption{Example shadow image (left) and the shadow subrings separated (right) with the intensity on a logarithmic scale. For the separated subrings, the top left is the $n=0$ subring that is the direct image of the disk, top right is $n=1$, bottom left is $n=2$, and bottom right is $n=3$. \label{fig:shadexamp}}
\end{figure*}

In the context of non-Kerr spacetimes and modified gravity theories, black hole shadows are usually treated in an idealistic manner,~i.e.~the photons that form the shadow image at the observing screen originate from spatial infinity isotropically around the black hole. While this certainly gives a projection of the photon sphere on the observing screen, it is not a realistic model for black hole shadow images. In reality, the black hole is surrounded by an accretion disk that is the source of the photons and any shadow image will be significantly affected by the geometry, astrophysics, and temporal evolution of the disk. In fact, the ideal shadow is not visible and instead the observed radiation is in the form of progressively thinner rings that approach the ideal shadow (assuming that the accretion disk is optically thin). In this work, the $n=0$ subring will refer to the direct image of the disk and each subsequent subring is made up of photons that make $n$ half-orbits of the black hole,~e.g.~the $n=1$ subring is made up of photons that crossed the equatorial plane of the black hole once before reaching the observer. Figure~\ref{fig:shadexamp} shows an example of the full shadow image and the subrings separated out, up to the $n=3$ subring (plotted on a log scale so the full ring is visible).

Recently, there has been much interest in the black hole shadow subrings as observables and what can be learned from them about the black hole spacetime. There have been a number of studies of the subrings in the Kerr spacetime (e.g.~\cite{Gralla:2019xty, Gralla:2019drh, Broderick:2021ohx, Bisnovatyi-Kogan:2022ujt}) with one of these including a deviation from the Kerr metric, albeit still within general relativity, through the inclusion of electric charge with the Kerr-Newman solution~\cite{Broderick:2021ohx}. Building off of some of the works with the Kerr solution, some others have repeated the studies with other non-Kerr black hole solutions (e.g.~\cite{Guo:2021bhr, Wielgus:2021peu, Gan:2021xdl}). A recent study using a more realistic disk model and non-Kerr solutions studied the black hole shadow as a whole, but found that the photon ring,~i.e.~where the higher order subrings stack up, is mostly unaffected by the astrophysics of the disk and can be used to extract spacetime parameters~\cite{Ozel:2021ayr, Younsi:2021dxe}. Overall, all of these works have confirmed that the shadow subrings, and particularly the higher order rings, are prime observables for studying the black hole spacetime and gravity.

This work attempts to expand on these past works by determining if and how well the parameters of the black hole spacetime and the accretion disk can be estimated both when analyzing the full image and the subrings individually. As part of this it is also studied how the parameter extraction is affected if the direct image,~i.e.~the $n=0$ subring, is removed, as has been proposed both through image processing techniques and making use of how the very long baseline interfometry (VLBI) of the EHT functions~\cite{Tiede:2020iif, Gelles:2021oyd}. Two different black hole spacetimes that parametrically deform the Kerr solution through various deformation parameters are studied, known as the Johannsen metric~\cite{Johannsen_2013} and the Konoplya-Rezzolla-Zhidenko (KRZ) metric~\cite{Konoplya_2016}. These are not necessarily solutions in any gravity theory, but  encompass a wide range of possible modifications to the Kerr metric. The accretion disk model used is a simple thin-disk model, but different emissivity profiles are included as a basic representation of how uncertainties in the disk physics can influence parameter extraction. A likelihood analysis is performed to determine if and how well the physical parameters can be estimated and the analysis is performed with two different resolutions of the image, $\sim 10~\mu$as and $\sim 1~\mu$as, to represent the resolutions of the current EHT and a future space-based version of the EHT, respectively.

The results are mixed in terms of what can be learned from black hole shadow images and subrings. At the $\sim 1~\mu$as resolution, the physical parameters can be recovered to a comparable or higher precision than what is possible today with other observations. This holds even if the $n=0$ subring is removed, in turn removing much of the uncertainty due to accretion disk physics, and suggests that a space-based version of the EHT would be extremely beneficial for understanding black holes and gravity. When focusing on the locations of the subrings at this higher resolution, the strong-gravity parameters of the spin and deformation from Kerr cannot be well-estimated, although the other parameters of the inclination angle, black hole mass, and the emissivity profile of the disk are usually well-estimated. At the lower resolution of $\sim 10~\mu$as very little can be said about the black hole spacetime when analyzing the location of the subrings, with poor estimates even for the mass and inclination angle. With the full image, one may be able to place weak constraints on the spin and deformation from Kerr, however these estimates are susceptible to bias from degeneracies between the parameters. It is possible that advanced image processing techniques and use of the specific advantages of VLBI can lead to better results than found here (e.g.~\cite{Tiede:2020iif, Gelles:2021oyd}), but one must certainly be careful about making any strong claims with the current capabilities of the EHT or any future earth-based VLBI telescopes.

The remainder of this paper is organized as follows: Sections~\ref{sec:spacetimes} and~\ref{sec:disk} describe the black hole spacetimes and the accretion disk model studied in this work, respectively. Section~\ref{sec:ray} details the ray-tracing code used to simulate the black hole shadow images. Section~\ref{sec:ana} presents the analysis methodology and the results are discussed in Section~\ref{sec:results}. Section~\ref{sec:conc} closes with a summary and brief discussion of this work. Throughout the work the metric signature $(-,+,+,+)$ and geometric units with $G=c=1$ are used, unless otherwise specified.

\section{Black Hole Spacetimes}
\label{sec:spacetimes}

In general relativity, the Kerr metric is the solution for an isolated, stationary, axisymmetric, and uncharged black hole. It is completely specified by just two physical parameters, the black hole mass $M$ and the black hole spin parameter $a\equiv J/M$, where $J\equiv|\vec{J}|$ is the magnitude of the black hole spin angular momentum. The line element of the Kerr metric in Boyer-Lindquist coordinates $(t,r,\theta,\phi)$ is
\begin{align}
ds^{2} = & -\left(1-\frac{2Mr}{\Sigma}\right)dt^{2} - \frac{4Mar\sin^{2}\theta}{\Sigma}dtd\phi + \frac{\Sigma}{\Delta}dr^{2}
\nonumber \\
& + \Sigma d\theta^{2} + \left(r^{2}+a^{2}+\frac{2Ma^{2}r\sin^{2}\theta}{\Sigma}\right)\sin^{2}\theta d\phi^{2},
\end{align}
where
\begin{equation}
\Sigma = r^{2}+a^{2}\cos^{2}\theta, \qquad \Delta = r^{2}-2Mr+a^{2}.
\end{equation}

In order to test the Kerr hypothesis and gravity, a non-Kerr black hole spacetime must be used (and preferably an analytic metric as it is more difficult to perform ray-tracing with a numerical metric). There are a number of black hole metrics in general relativity and modified gravity theories that are non-Kerr (e.g.~\cite{Maselli_2015, Bambi_2017}), however the number of solutions makes it difficult to choose those that are most representative of possible departures from the Kerr metric and general relativity. For that reason, two parameterized metrics,~i.e.~metrics that encode departures from the Kerr metric through one or more parameters, are used in this work: the Johannsen metric~\cite{Johannsen_2013} and the Konoplya-Rezzolla-Zhidenko metric~\cite{Konoplya_2016}. The Johannsen metric introduces parameterized modifications to the Kerr metric that are motivated by a parameterized post-Newtonian expansion, which is one way to encode departures from general relativity. Note that the Johannsen metric is a subset of the wider class of modified gravity metrics proposed by Vigeland, Yunes, and Stein~\cite{Vigeland_2011}. The KRZ metric uses a continued fraction expansion for its deformation functions, which should have improved convergence properties over the usual expansion in powers of $M/r$ as well as being better able to represent near-horizon modifications. Additionally, unlike the Johannsen metric, the KRZ metric does not necessarily possess a Carter-like constant and so has a different symmetry structure from Kerr, which allows it to represent a different set of departures from Kerr.

\subsection{Johannsen Metric}
\label{sec:jo}

The Johannsen metric~\cite{Johannsen_2013} is stationary, axisymmetric, and asymptotically flat, as well as having a Carter-like constant, so it has similar symmetry properties to the Kerr metric. The line element in Boyer-Lindquist coordinates $(t,r,\theta,\phi)$ is given by
\begin{align}
ds^{2} = & -\frac{-\tilde{\Sigma}\left(\Delta-a^{2}A_{2}^{2}\sin^{2}\theta\right)}{B^{2}} dt^{2}+\frac{\tilde{\Sigma}}{\Delta A_{5}}dr^{2}+\tilde{\Sigma} d\theta^{2}
\nonumber \\
& +\frac{\left[\left(r^{2}+a^{2}\right)^{2}A_{1}^{2}-a^{2}\Delta\sin^{2}\theta\right]\tilde{\Sigma}\sin^{2}\theta}{B^{2}}d\phi^{2}
\nonumber \\
& -\frac{2a\left[\left(r^{2}+a^{2}\right)A_{1}A_{2}-\Delta\right]\tilde{\Sigma}\sin^{2}\theta}{B^{2}}dtd\phi,
\end{align}
where
\begin{align}
B = & \left(r^{2}+a^{2}\right)A_{1}-a^{2}A_{2}\sin^{2}\theta, \quad \tilde{\Sigma} = \Sigma+f,
\nonumber \\
\Sigma = & r^{2}+a^{2}\cos^{2}\theta, \quad \Delta = r^{2}-2Mr+a^{2}, \label{eq-sigdel}
\end{align}
the four free functions $f$, $A_{1}$, $A_{2}$, and $A_{5}$, are
\begin{align}
f = & \sum_{n=3}^{\infty}\sigma_{n}\frac{M^{n}}{r^{n-2}}, \quad A_{1} = 1+\sum_{n=3}^{\infty}\alpha_{1n}\left(\frac{M}{r}\right)^{n},
\nonumber \\
A_{2} = & 1+\sum_{n=2}^{\infty}\alpha_{2n}\left(\frac{M}{r}\right)^{n}, \quad A_{5} = 1+\sum_{n=2}^{\infty}\alpha_{5n}\left(\frac{M}{r}\right)^{n}.
\end{align}
The four free functions have additional lower-order terms that can either be absorbed into the definition of $M$ and $a$ or vanish to satisfy Solar System constraints (see~\cite{Johannsen_2013} for more details). Note that when $\epsilon_{n}=\alpha_{1n}=\alpha_{2n}=\alpha_{5n}=0$ the metric reduces to the Kerr metric. There is a more general extension of the Johannsen metric that was recently found and studied~\cite{Carson_2020}, however the new modifications in this extension require multiple non-zero deformation parameters to prevent the appearance of a naked singularity, and so this metric is not used in this work.

In this work, only the leading-order deformation parameters will be studied and they will be studied individually,~i.e.~with all but one parameter set to zero at a time, as it is quite computationally time-consuming to study even just two non-zero deformation parameters simultaneously. Then, the version of the Johannsen metric used in this work, with the deformation parameters shown explicitly, is
\begin{align}
ds^{2} = & -\frac{\left(\Sigma+\epsilon_{3}\frac{M^{3}}{r}\right)\left(\Delta-a^{2}\left(1+\alpha_{22}\left(\frac{M}{r}\right)^{2}\right)^{2}\sin^{2}\theta\right)}{\left(r^{2}+a^{2}\right)\left(1+\alpha_{13}\left(\frac{M}{r}\right)^{3}\right)-a^{2}\left(1+\alpha_{22}\left(\frac{M}{r}\right)^{2}\sin^{2}\theta\right)} dt^{2}
\nonumber \\
& +\frac{\left(\Sigma+\epsilon_{3}\frac{M^{3}}{r}\right)}{\Delta\left(1+\alpha_{52}\left(\frac{M}{r}\right)^{2}\right)}dr^{2}+\left(\Sigma+\epsilon_{3}\frac{M^{3}}{r}\right) d\theta^{2}
\nonumber \\
& +\frac{\left[\left(r^{2}+a^{2}\right)^{2}\left(1+\alpha_{13}\left(\frac{M}{r}\right)^{3}\right)^{2}-a^{2}\Delta\sin^{2}\theta\right]\left(\Sigma+\epsilon_{3}\frac{M^{3}}{r}\right)\sin^{2}\theta}{\left(r^{2}+a^{2}\right)\left(1+\alpha_{13}\left(\frac{M}{r}\right)^{3}\right)-a^{2}\left(1+\alpha_{22}\left(\frac{M}{r}\right)^{2}\sin^{2}\theta\right)}d\phi^{2}
\nonumber \\
& -\frac{2a\left[\left(r^{2}+a^{2}\right)\left(1+\alpha_{13}\left(\frac{M}{r}\right)^{3}\right)\left(1+\alpha_{22}\left(\frac{M}{r}\right)^{2}\right)-\Delta\right]\left(\Sigma+\epsilon_{3}\frac{M^{3}}{r}\right)\sin^{2}\theta}{\left(r^{2}+a^{2}\right)\left(1+\alpha_{13}\left(\frac{M}{r}\right)^{3}\right)-a^{2}\left(1+\alpha_{22}\left(\frac{M}{r}\right)^{2}\sin^{2}\theta\right)}dtd\phi.
\end{align}

Additionally, the conditions that there is an event horizon with no naked singularity and that there are no numerical/physical pathologies in the spacetime are imposed. For the Kerr spacetime, this is easily guaranteed by imposing that $a/M\leq1$ so that there exists an event horizon. The Johannsen spacetime has the same condition, but also has conditions on the deformation parameters to avoid pathologies such as closed time-like curves or violations of the Lorentzian signature outside of the event horizon. This is imposed by requiring that outside of the event horizon the metric determinant is negative, the metric element $g_{\phi\phi}>0$, and the metric does not diverge. These lead to the following constraints on the deformation parameters
\begin{align}
\alpha_{13} > & -\frac{1}{2}\left(1+\sqrt{1-\chi^{2}}\right)^4, \label{eq-a13bound}
\\
-\left(1+\sqrt{1-\chi^{2}}\right)^{2} & < \alpha_{22} < \frac{\left(1+\sqrt{1-\chi^{2}}\right)^{4}}{\chi^{2}}, \label{eq-a22bound}
\end{align}
where $\chi\equiv a/M$ is the dimensionless spin parameter.

\subsection{KRZ Metric}
\label{sec:krz}

Like the Johannsen metric, the KRZ metric~\cite{Konoplya_2016} is stationary, axisymmetric, and asymptotically flat, however it does not necessarily possess a Carter-like constant and so does not have similar symmetry properties to the Kerr metric. The line element in Boyer-Lindquist coordinates $(t,r,\theta,\phi)$ is
\begin{equation}
ds^{2} = -\frac{N^{2}-W^{2}\sin^{2}\theta}{K^{2}}dt^{2} +\frac{\Sigma B^{2}}{N^{2}}dr^{2} +\Sigma r^{2}d\theta^{2} +K^{2}r^{2}\sin^{2}\theta d\phi^{2} -2Wr\sin^{2}\theta dtd\phi,
\end{equation}
where
\begin{align}
N^{2} = & xA_{0}+\sum_{i=1}^{\infty}A_{i}y^{i}, \qquad B = 1+\sum_{i=1}^{\infty}B_{i}y^{i},
\nonumber \\
W = & \sum_{i=0}^{\infty}\frac{w_{i}y^{i}}{\Sigma}, \qquad K^{2} = 1 + \frac{aW}{r} + \sum_{i=0}^{\infty}\frac{k_{i}y^{i}}{\Sigma},
\end{align}
and
\begin{align}
B_{i} = & b_{i0}\left(1-x\right)+\tilde B_{i}\left(1-x\right)^{2},
\\
W_{i} = & w_{i0}\left(1-x\right)^{2}+\tilde W_{i}\left(1-x\right)^{3},
\\
K_{i} = & k_{i0}\left(1-x\right)^{2}+\tilde K_{i}\left(1-x\right)^{3},
\\
A_{0} = & 1-\epsilon_{0}\left(1-x\right)+\left(a_{00}-\epsilon_{0}+k_{00}\right)\left(1-x\right)^{2}+\tilde A_{0}\left(1-x\right)^{3},
\\
A_{i>0} = & k_{i}+\epsilon_{i}\left(1-x\right)^{2}+a_{i0}\left(1-x\right)^{3}+\tilde A_{i}\left(1-x\right)^{4},
\\
\Sigma = & 1+\frac{a^{2}}{r_{0}^{2}}\left(1-x\right)^{2}y^{2}.
\end{align}
The continued fraction expansion that is the main feature of the KRZ metric shows up in the four functions
\begin{align}
\tilde A_{i} = & \frac{a_{i1}}{1+\frac{a_{i2}x}{1+\frac{a_{i3}x}{1+...}}}, \qquad \tilde B_{i} = \frac{b_{i1}}{1+\frac{b_{i2}x}{1+\frac{b_{i3}x}{1+...}}},
\nonumber \\
\tilde W_{i} = & \frac{w_{i1}}{1+\frac{w_{i2}x}{1+\frac{w_{i3}x}{1+...}}}, \qquad \tilde K_{i} = \frac{k_{i1}}{1+\frac{k_{i2}x}{1+\frac{k_{i3}x}{1+...}}}.
\end{align}
Here, $x=1-r_{0}/r$, $y=\cos\theta$, and $r_{0}$ is the horizon radius in the equatorial plane that must be found by solving for the largest solution of $N^{2}\left(r,\pi/2\right)=0$. The coefficients $\epsilon_{i},a_{ij},b_{ij},w_{ij},k_{ij}$ for $i=0,1,2,3,...,j=1,2,3,...$ are the parameters of the metric along with the spin parameter $a$. The black hole mass $M$, in principle, appears in the expression for the equatorial horizon radius $r_{0}$. Note that the coefficients $\epsilon_{i},a_{i0},b_{i0},w_{i0},k_{i0}$ for $i=0,1,2,3,...$ are determined by the desired asymptotic behavior near spatial infinity.

For the purposes of this work, only the leading order coefficients will be considered. In that case the metric functions are
\begin{align}
N^{2} = & \left(1-\frac{r_0}{r}\right)\left[1-\frac{\epsilon_{0}r_{0}}{r}+\left(k_{00}-\epsilon_{0}\right)\frac{r_{0}^{2}}{r^{2}}+\frac{\delta_{1}r_{0}^{3}}{r^{3}}\right]
\nonumber \\
&+\left[\left(k_{21}+a_{20}\right)\frac{r_{0}^{3}}{r^{3}}+\frac{a_{21}r_{0}^{4}}{r^{4}}\right]\cos^{2}\theta,
\\
B = & 1+\frac{\delta_{4}r_{0}^{2}}{r^{2}}+\frac{\delta_{5}r_{0}^{2}}{r^{2}}\cos^{2}\theta,
\\
W = & \frac{1}{\Sigma}\left(\frac{w_{00}r_{0}^{2}}{r^{2}}+\frac{\delta_{2}r_{0}^{3}}{r^{3}}+\frac{\delta_{3}r_{0}^{3}}{r^{3}}\cos^{2}\theta\right),
\\
K^{2} = & 1+\frac{\chi W}{r}+\frac{1}{\Sigma}\left(\frac{k_{00}r_{0}^{2}}{r^{2}}+\frac{k_{21}r_{0}^{3}}{r^{3}}\cos^{2}\theta\right),
\\
\Sigma = & 1+\frac{\chi^{2}}{r^{2}}\cos^{2}\theta,
\end{align}
where, for convenience, the six deformation parameters $\delta_{j}$ for $j=1,2,...,6$ have been introduced and are related to the KRZ coefficients by
\begin{align}
a_{20} = & \frac{2\chi^{2}}{r_{0}^{3}}, \qquad a_{21} = -\frac{\chi^{4}}{r_{0}^{4}}+\delta_{6}, \qquad \epsilon_{0} = \frac{2-r_{0}}{r_{0}},
\nonumber \\
k_{00} = & \frac{\chi^{2}}{r_{0}^{2}}, \qquad k_{21} = \frac{\chi^{4}}{r_{0}^{4}}-\frac{2\chi^{2}}{r_{0}^{3}}-\delta_{6}, \qquad w_{00} = \frac{2\chi}{r_{0}^{2}},
\end{align}
and the horizon radius matches that of Kerr, $r_{0} = 1+\sqrt{1-\chi^{2}}$. Note that here the black hole mass has been set to unity for simplicity, $M=1$.

As in the case of the Johannsen spacetime, in order to exclude pathologies in the spacetime and require an event horizon with no naked singularity the following conditions on the deformation parameters for the KRZ spacetime are imposed
\begin{align}
\delta_{1} & > \frac{4r_{0}-3r_{0}^{2}-\chi^{2}}{r_{0}^{2}}, \label{eq-d1bound}
\\
\delta_{2},\delta_{3} & \left\{\begin{array}{l}
		> \\
		< \\
		\end{array} -\frac{4}{a^3}(1-\sqrt{1-a^2}) \quad \begin{array}{l}
		 \rm{if} \; a > 0\\ 
		 \rm{if} \; a < 0,
		\end{array} \right. \label{eq-d23bound}
\\
\delta_{4},\delta_{5} & > -1, \label{eq-d45bound}
\\
\delta_{6} & < \frac{r_{0}^{2}}{4-\chi^{2}}. \label{eq-d6bound}
\end{align}

\subsection{Event Horizon and ISCO}
\label{sec:ISCO}

For this work it is necessary to determine the radius of the event horizon and the innermost stable circular orbit (ISCO) for massive particles within the Johannsen and KRZ spacetimes. The event horizon radius defines the point of no return for a photon, and so any numerical calculations do not need to proceed once this radius is crossed (in reality the calculation is stopped just outside the event horizon as it is a coordinate singularity in Boyer-Lindquist coordinates). The ISCO radius defines the inner point of the standard Novikov-Thorne thin disk model ~\cite{1973blho.conf..343N}, and in this work will set the boundary between the standard accretion disk with particles on quasi-Keplerian orbits and the inner disk with particles that are on plunging orbits into the event horizon.

An event horizon can be defined as a null surface generated by null geodesic generators,~i.e.~the surface at which radially outgoing photons are confined. The normal to a null surface $n^{\mu}$ must itself be null, so the event horizon must satisfy the condition
\begin{equation}
g^{\mu\nu}\partial_{\mu}F\partial_{\nu}F=0,
\end{equation}
where $F(x^{\alpha})$ is a level surface function such that $n_{\mu}=\partial_{\mu}F$. The Kerr, Johannsen, and KRZ spacetimes are stationary, axisymmetric, and reflection symmetric about the poles and equator, and so the level surfaces can only depend on radius. Then, letting $F(x^{\alpha})=r-r_{\text{H}}$, where $F=0$ defines the location of the event horizon, simplifies the above condition to $g^{rr}=0=g_{tt}g_{\phi\phi}-g_{t\phi}^{2}$~\cite{2004rtmb.book.....P}. For both the Johannsen spacetime and the KRZ spacetime the horizon radius matches that of the Kerr spacetime
\begin{equation}
r_{\text{H}}=M+\sqrt{M-a^{2}}. \label{eq:horizon}
\end{equation}

\begin{figure*}[hpt]
\includegraphics[width=0.33\columnwidth{},clip=true]{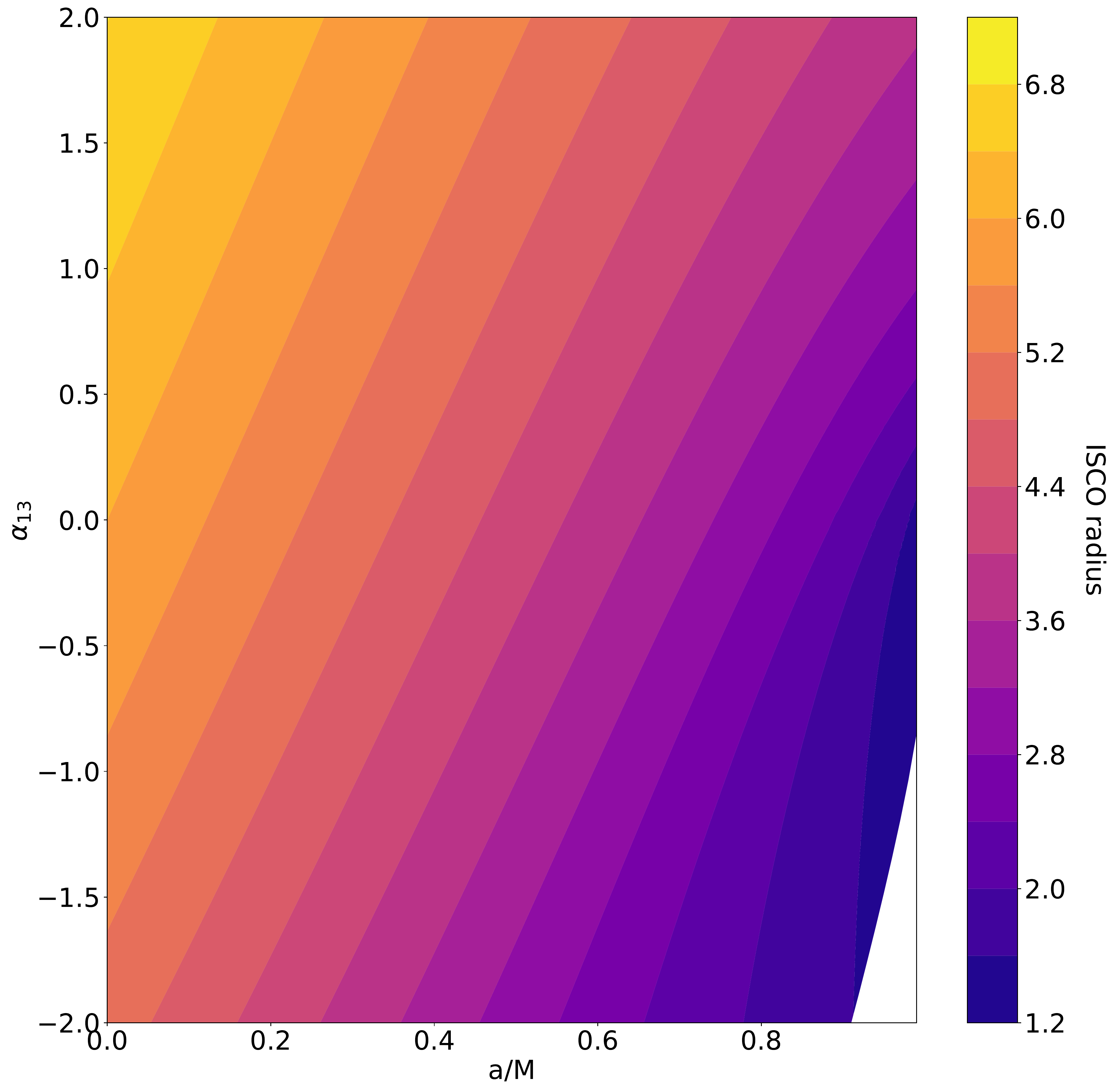}
\includegraphics[width=0.33\columnwidth{},clip=true]{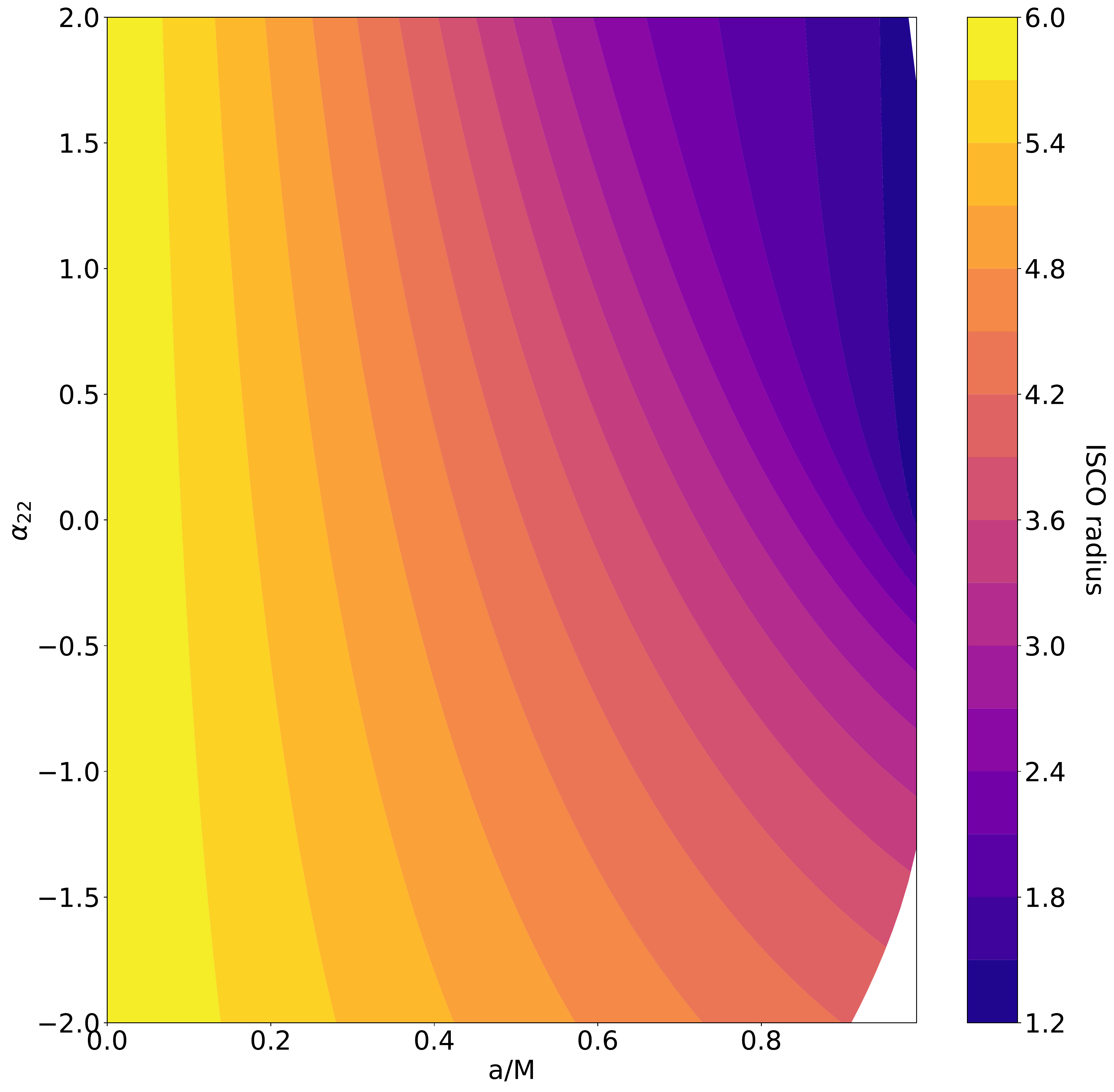}
\includegraphics[width=0.33\columnwidth{},clip=true]{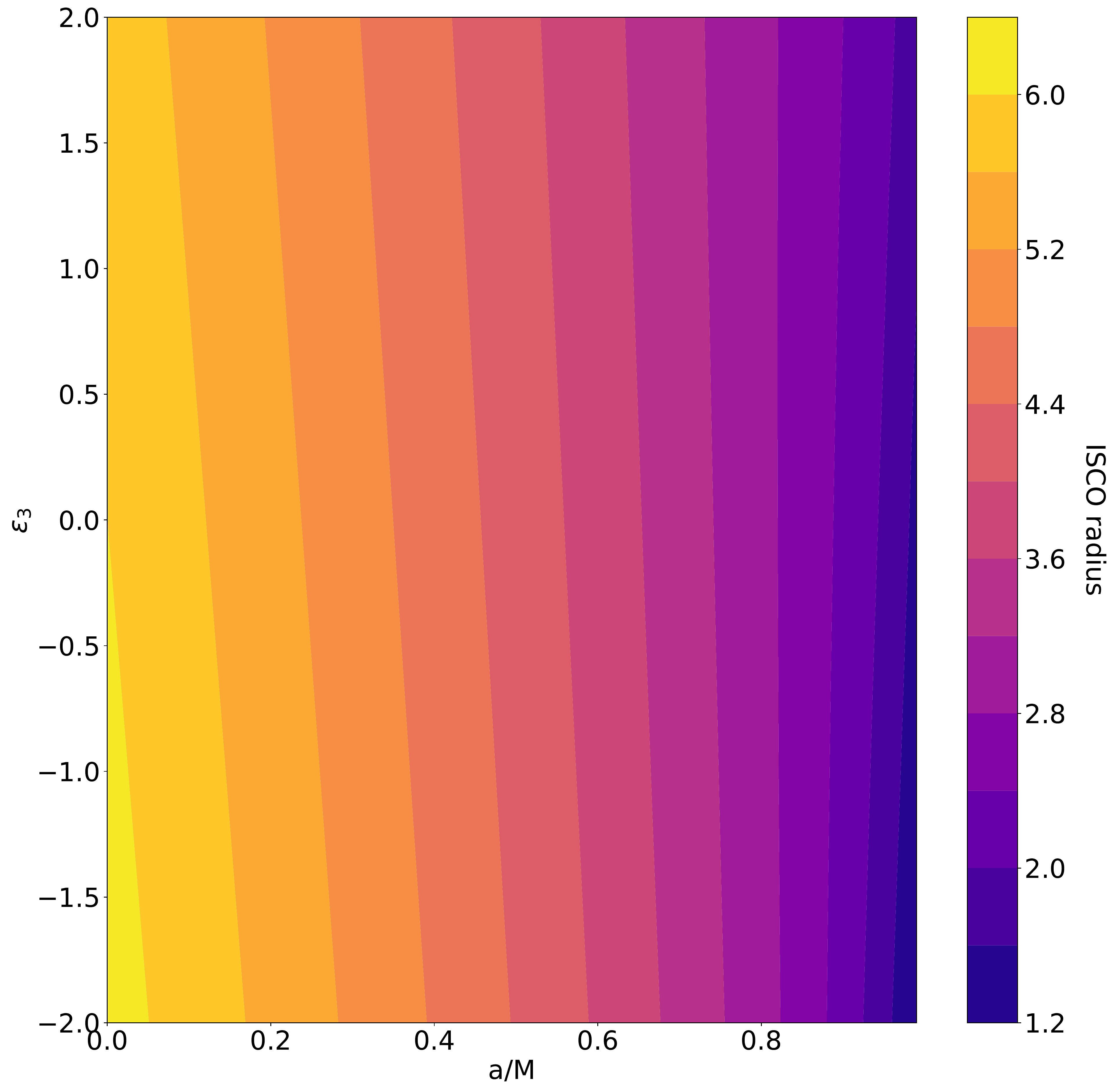}
\caption{ISCO radius as a function of spin and deformation parameter in the Johannsen spacetime. Only one deformation parameter is non-zero in each plot ($\alpha_{13}$, $\alpha_{22}$, and $\epsilon_{3}$, from left to right). The white regions on the plots are those that are excluded to avoid pathologies (see~Eqs.~\ref{eq-a13bound} and~\ref{eq-a22bound}). Note that for the $\alpha_{52}$ case the ISCO radius is independent of $\alpha_{52}$ and matches that of Kerr. \label{fig:iscoJo}}
\end{figure*}

\begin{figure*}[hpt]
\includegraphics[width=0.5\columnwidth{},clip=true]{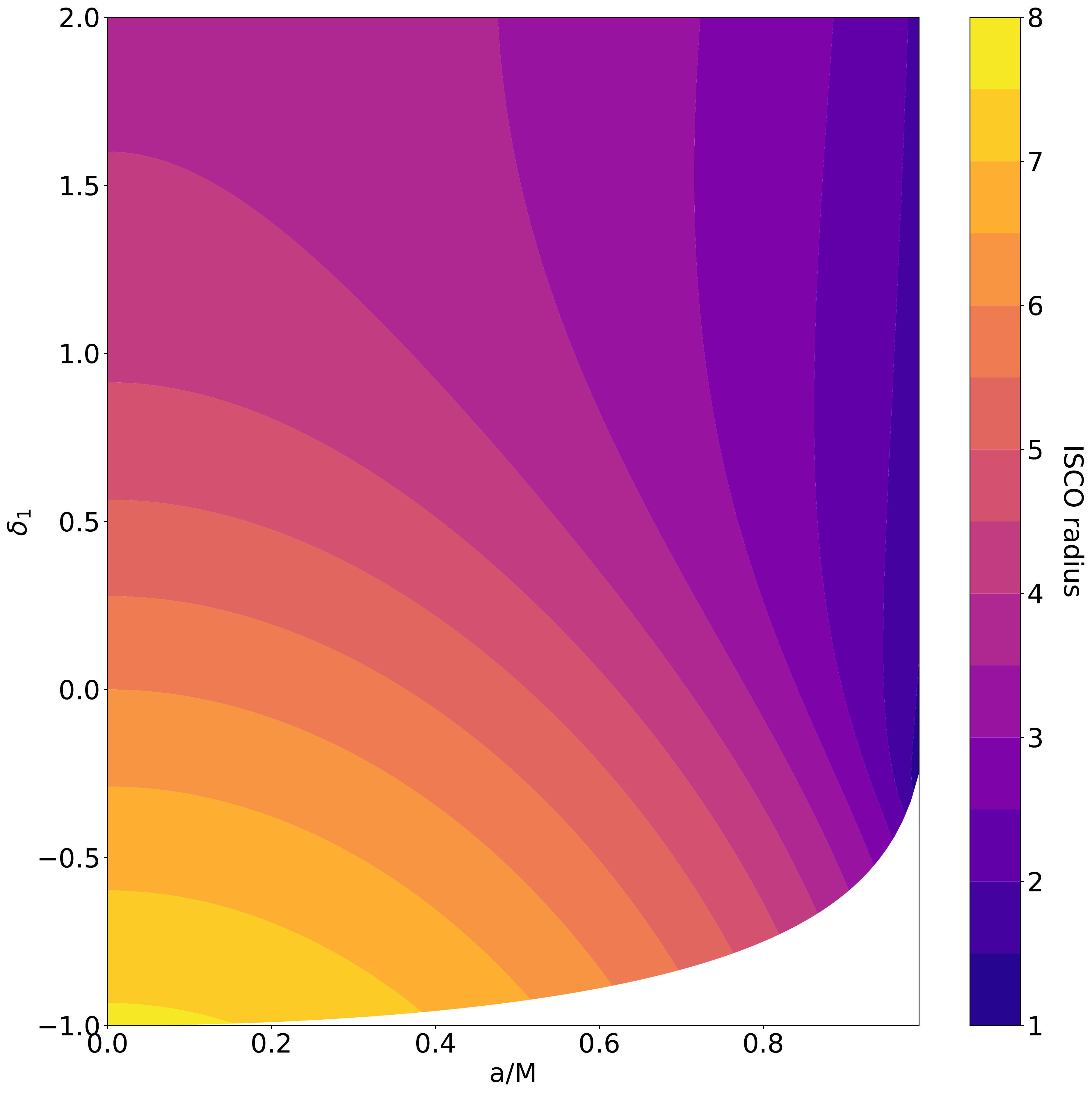}
\includegraphics[width=0.5\columnwidth{},clip=true]{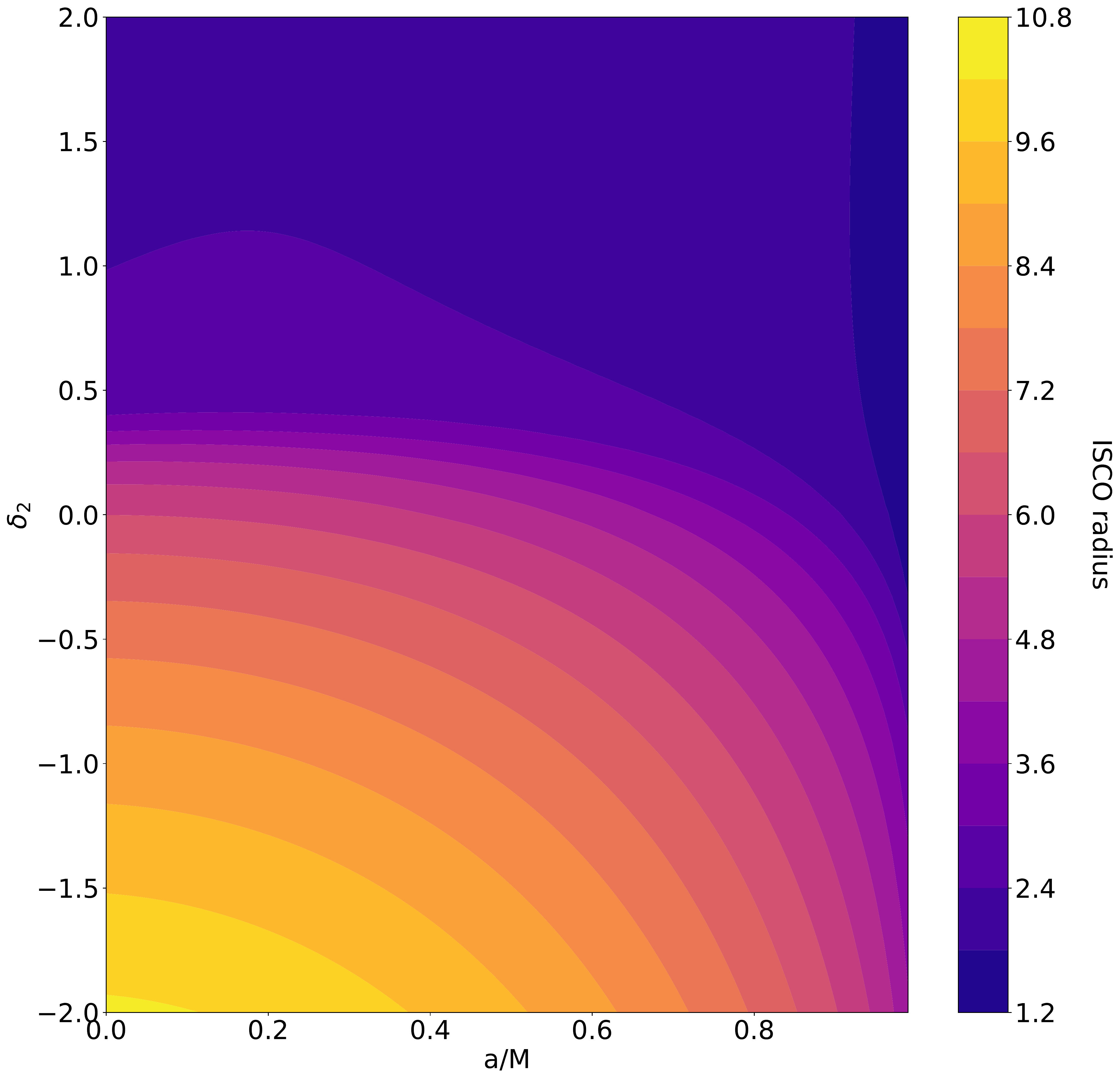}
\caption{ISCO radius as a function of spin and deformation parameter in the KRZ spacetime. Only one deformation parameter is non-zero in each plot ($\delta_{1}$ on the left and $\delta_{2}$ on the right). The white region on the plot is that excluded to avoid pathologies (see~Eq.~\ref{eq-d1bound}). Note that for the $\delta_{3}$, $\delta_{4}$, $\delta_{5}$, and $\delta_{6}$ cases the ISCO radius is independent of the deformation parameter and matches that of Kerr. \label{fig:iscoKRZ}}
\end{figure*}

As the Johannsen and KRZ spacetimes are stationary and axisymmetric, they contain a timelike Killing vector and an azimuthal Killing vector. These Killing vectors imply the existence of two conserved quantities, the energy and the ($z$-component of the) angular momentum. From the definitions of $E$ and $L_{z}$ one can write
\begin{align}
\dot{t} = & \frac{Eg_{\phi\phi}+L_{z}g_{t\phi}}{g_{t\phi}^{2}-g_{tt}g_{\phi\phi}}, \label{eq-dott}
\\
\dot{\phi} = & -\frac{Eg_{t\phi}+L_{z}g_{tt}}{g_{t\phi}^{2}-g_{tt}g_{\phi\phi}}, \label{eq-dotphi}
\end{align}
where the overhead dot represents a derivative with respect to an affine parameter (proper time for a massive particle). Substituting into the normalization condition for the four-velocity of massive particles, $u^{\mu}u_{\mu}=-1$, one finds
\begin{equation}
g_{rr}\dot{r}^{2}+g_{\theta\theta}\dot{\theta}^{2}=V_{\text{eff}}\left(r,\theta;E,L_{z}\right),
\end{equation}
where the effective potential is
\begin{equation}
V_{\text{eff}}\equiv\frac{E^{2}g_{\phi\phi}+2EL_{z}g_{t\phi}+L_{z}^{2}g_{tt}}{g_{t\phi}^{2}-g_{tt}g_{\phi\phi}}-1. \label{eq:veff}
\end{equation}
By imposing equatorial circular orbits, explicit expressions for the energy and angular momentum can be found. Requiring circularity is equivalent to requiring that $V_{\text{eff}}=0=\partial V_{\text{eff}}/\partial r$, and solving these for $E$ and $L_{z}$, the expressions are
\begin{align}
E = & -\frac{g_{tt}+g_{t\phi}\Omega}{\sqrt{-\left(g_{tt}+2g_{t\phi}\Omega+g_{\phi\phi}\Omega^{2}\right)}}, \label{eq-E}
\\
L_{z} = & \frac{g_{t\phi}+g_{\phi\phi}\Omega}{\sqrt{-\left(g_{tt}+2g_{t\phi}\Omega+g_{\phi\phi}\Omega^{2}\right)}}, \label{eq-Lz}
\end{align}
where the angular velocity of equatorial circular geodesics is
\begin{equation}
\Omega=\frac{d\phi}{dt}=\frac{-g_{t\phi,r}\pm\sqrt{\left(g_{t\phi,r}\right)^2-g_{tt,r}g_{\phi\phi,r}}}{g_{\phi\phi,r}}. \label{eq-omega}
\end{equation}
The ISCO radius can be found by substituting the expressions for $E$ and $L_{z}$ into the expression for $V_{\text{eff}}$ (Eq.~\ref{eq:veff}) and solving $\partial^{2}V_{\text{eff}}/\partial r^{2}=0$ for $r$. In Kerr, the ISCO radius is given by
\begin{equation}
r_{\text{ISCO}}=M\left(3+Z_{2}-\left[\left(3-Z_{1}\right)\left(3+Z_{1}+2Z_{2}\right)\right]^{1/2}\right),
\end{equation}
where
\begin{align}
Z_{1} = & 1+\left(1-\chi^{2}\right)^{1/3}\left[\left(1+\chi\right)^{1/3}+\left(1-\chi\right)^{1/3}\right],
\\
Z_{2} = & \left(3\chi^{2}+Z_{1}^{2}\right)^{1/2}.
\end{align}
The ISCO radius for the Johannsen spacetime is independent of $\alpha_{52}$ and for the KRZ spacetime is independent of $\delta_{3}$, $\delta_{4}$, $\delta_{5}$, and $\delta_{6}$, and so for those cases the ISCO radius matches that of Kerr. Otherwise, there are not nice expressions for the ISCO radius of the Johannsen and KRZ spacetimes. Instead, see Figs.~\ref{fig:iscoJo} and~\ref{fig:iscoKRZ} for plots of the ISCO radius as a function of spin and deformation parameter.

%

\section{Accretion Disk Model}
\label{sec:disk}

The source of radiation for the black hole shadow of a supermassive black hole is the accretion disk that is around the black hole. Accretion disks are quite complex physical systems and require detailed general relativistic magnetohydrodynamic (GRMHD) simulations to model them properly (and even then there are a number of open questions about the physics that goes into the modeling). For the purposes of this work a simple accretion disk model will be used, however with different emissivity profiles to represent some of the lack of knowledge of the correct model. The accretion disk is modeled along the lines of the geometrically thin Novikov-Thorne model~\cite{1973blho.conf..343N}. While the Novikov-Thorne model is generally treated as optically thick, here the disk will be treated as optically thin to allow for higher order images of the disk that may pass through the disk to be visible to the observer. This is in agreement with what is known of the disk surrounding the black hole at the center of M87 as observed by the EHT~\cite{2019ApJ...875L...1E}. Additionally, the emission from the disk will be treated achromatically,~i.e.~independent of frequency. While redshift will be calculated as part of calculating the radiation intensity, the actual emitted and observed frequency of the radiation will be ignored. This is obviously not the case in reality, and in fact the regions of the shadow image that are dependent on the disk physics have a fairly significant frequency dependence. These simplifications are made as part of this first look exploration of testing the Kerr solution and gravity with shadow subrings and a future, more in-depth study would be needed to include a more realistic disk model.

As with the Novikov-Thorne model, the disk is confined to the equatorial plane of the black hole spacetime and is treated as infinitesimally thin,~i.e.~$\theta=\pi/2$ and $\dot\theta=0$, where the overhead dot represents a derivative with respect to the proper time. Usually in the Novikov-Thorne model the accretion disk terminates at the ISCO radius as the matter below the ISCO radius is thought to very quickly plunge into the black hole. In the model used here, the plunging gas and the radiation it emits is included. To do so, the disk is split into two regions: the standard thin disk that ends at the ISCO radius and the plunging gas that goes from the ISCO to the event horizon.

\subsection{Thin Disk}

The expressions for the energy and angular momentum of massive particles in the disk have already been derived in Sec.~\ref{sec:ISCO} and are given by Eqs.~\ref{eq-E} and~\ref{eq-Lz}.

In order to calculate the radiation intensity as seen by the observer, one must first calculate the redshift of the photons emitted from the disk and observed by some observer at spatial infinity. The redshift can be defined by
\begin{equation}
g\equiv\frac{\nu_{o}}{\nu_{e}}=\frac{p_{\mu}u_{o}^{\mu}}{p_{\nu}u_{e}^{\nu}}, \label{eq-redshift}
\end{equation}
where $\nu_{o}$ and $\nu_{e}$ are the photon frequency as measured at the observer and the emitter, respectively, $p_{\mu}$ is the photon canonical conjugate momentum, and $u_{o}^{\mu}$ and $u_{e}^{\mu}$ are the four-velocities of the observer and the emitting material in the disk, respectively.

The spacetimes studied here are stationary and axisymmetric, so the photon's conjugate momentum can be written as $p_{\mu}=\left(-E^{\gamma},p_{r}^{\gamma},p_{\theta}^{\gamma},L_{z}^{\gamma}\right)$. The observer can be treated as static, $u_{o}^{\mu}=\left(1,0,0,0\right)$, and the four velocity of the orbiting material is as calculated earlier in this section, $u_{e}^{\mu}=u_{e}^{t}\left(1,0,0,\Omega\right)$. Equation~\ref{eq-omega} gives $\Omega$ and $u_{e}^{t}=\dot t$ can be found by substituting Eqs.~\ref{eq-E} and~\ref{eq-Lz} into Eq.~\ref{eq-dott},
\begin{equation}
\dot t = \frac{1}{\sqrt{-\left(g_{tt}+2g_{t\phi}\Omega+g_{\phi\phi}\Omega^{2}\right)}}.
\end{equation}
The redshift of photons emitted from the thin disk is then
\begin{equation}
g=\frac{\sqrt{-\left(g_{tt}+2g_{t\phi}\Omega+g_{\phi\phi}\Omega^{2}\right)}}{1-b\Omega},
\end{equation}
where $b\equiv L_{z}^{\gamma}/E^{\gamma}$ is a conserved quantity of the photon trajectory sometimes referred to as the impact parameter.

\subsection{Plunging region}

In order to calculate the redshift of photons emitted from the gas in the plunging region the gas in the disk is assumed to follow quasi-circular orbits throughout the thin disk region, down to the ISCO radius, below which the gas follows infalling geodesics down to the event horizon. Thus, the energy and angular momentum of the gas is that which it has at the ISCO radius, effectively assuming there is no longer energy and angular momentum transfer in the plunging region. The ensuing calculation follows that performed in~\cite{C_rdenas_Avenda_o_2020}, however here it is done for a general stationary axisymmetric black hole spacetime rather than specializing to the Kerr spacetime.

The redshift is still defined by Eq.~\ref{eq-redshift} and the four-velocity of the observer is unchanged, but the four-velocity of the emitting material is now that of plunging material,~i.e.~with a radial component, $u_{e,p}^{\mu}=\left(u_{e,p}^{t},u_{e,p}^{r},0,u_{e,p}^{\phi}\right)$. The $t-$ and $\phi-$ components of the four-velocity come from the energy and angular momentum and are given by Eqs.~\ref{eq-dott} and~\ref{eq-dotphi}, however with $E$ and $L_{z}$ set to their values for a circular orbit at the ISCO radius. For completeness, they are
\begin{align}
u_{e,p}^{t}=-\frac{E_{\text{ISCO}}g_{\phi\phi}+L_{\text{ISCO}}g_{t\phi}}{g_{tt}g_{\phi\phi}-g_{t\phi}^{2}},
\\
u_{e,p}^{\phi}=\frac{E_{\text{ISCO}}g_{t\phi}+L_{\text{ISCO}}g_{tt}}{g_{tt}g_{\phi\phi}-g_{t\phi}^{2}},
\end{align}
where
\begin{align}
E_{\text{ISCO}} = & \left. -\frac{g_{tt}+g_{t\phi}\Omega_{\text{ISCO}}}{\sqrt{-\left(g_{tt}+2g_{t\phi}\Omega_{\text{ISCO}}+g_{\phi\phi}\Omega_{\text{ISCO}^{2}}\right)}}\right\rvert_{r=r_{\text{ISCO}}},
\\
L_{\text{ISCO}} = & \left. \frac{g_{t\phi}+g_{\phi\phi}\Omega}{\sqrt{-\left(g_{tt}+2g_{t\phi}\Omega_{\text{ISCO}}+g_{\phi\phi}\Omega_{\text{ISCO}^{2}}\right)}}\right\rvert_{r=r_{\text{ISCO}}},
\end{align}
and
\begin{align}
\Omega_{\text{ISCO}} = \left.\frac{-g_{t\phi,r}+\sqrt{\left(g_{t\phi,r}\right)^{2}-g_{tt,r}g_{\phi\phi,r}}}{g_{\phi\phi,r}}\right\rvert_{r=r_{\text{ISCO}}}.
\end{align}
The radial component can then by calculated through the normalization condition $u^{\mu}u_{\mu}=-1$,
\begin{equation}
u_{e,p}^{r}=-\sqrt{\frac{-\left(1+g_{tt}\left(u_{e,p}^{t}\right)^{2}+g_{\phi\phi}\left(u_{e,p}^{\phi}\right)^{2}+2g_{t\phi}u_{e,p}^{t}u_{e,p}^{\phi}\right)}{g_{rr}}}.
\end{equation}
Note that the metric terms in the four-velocity components are not calculated at the ISCO radius, only those that are in the expressions for the energy, angular momentum, and angular velocity.

With the four-velocity of the plunging gas calculated, the redshift of photons emitted from the plunging gas and observed at spatial infinity is
\begin{equation}
g_{p}=\frac{1}{u_{e,p}^{t}-bu_{e,p}^{\phi}-p^{r}_{\gamma}g_{rr}u_{e,p}^{r}},
\end{equation}
where by conservation of energy it has been used that $-E^{\gamma}=p_{t}^{\gamma}=g_{tt}p_{\gamma}^{t}=-p^{t}_{\gamma,o}\approx-1$ at the spatial infinity of the observer.

\subsection{Emissivity Profile and Intensity}
\label{sec:emprof}

In addition to the redshift, calculating the black hole shadow requires the emissivity profile of the accretion disk,~i.e.~the relation between the radius and the intensity of radiation emitted from a location in the disk (the relation is only radially dependent here due to the axial symmetry of the system). The emissivity profile of the disk is strongly dependent on the astrophysical properties of the disk and can be quite complex if the disk is similarly complex.

For simplicity and since a simple disk model is used throughout this work, a pair of similarly simple emissivity profiles will be used. The profile within the thin disk region will be that of a power-law
\begin{equation}
\epsilon \propto \left(\frac{r_{\text{ISCO}}}{r}\right)^{q}, \label{eq-emisd}
\end{equation}
where $q$ is the emissivity index.

\begin{figure*}[hpt]
\includegraphics[width=\columnwidth{},clip=true]{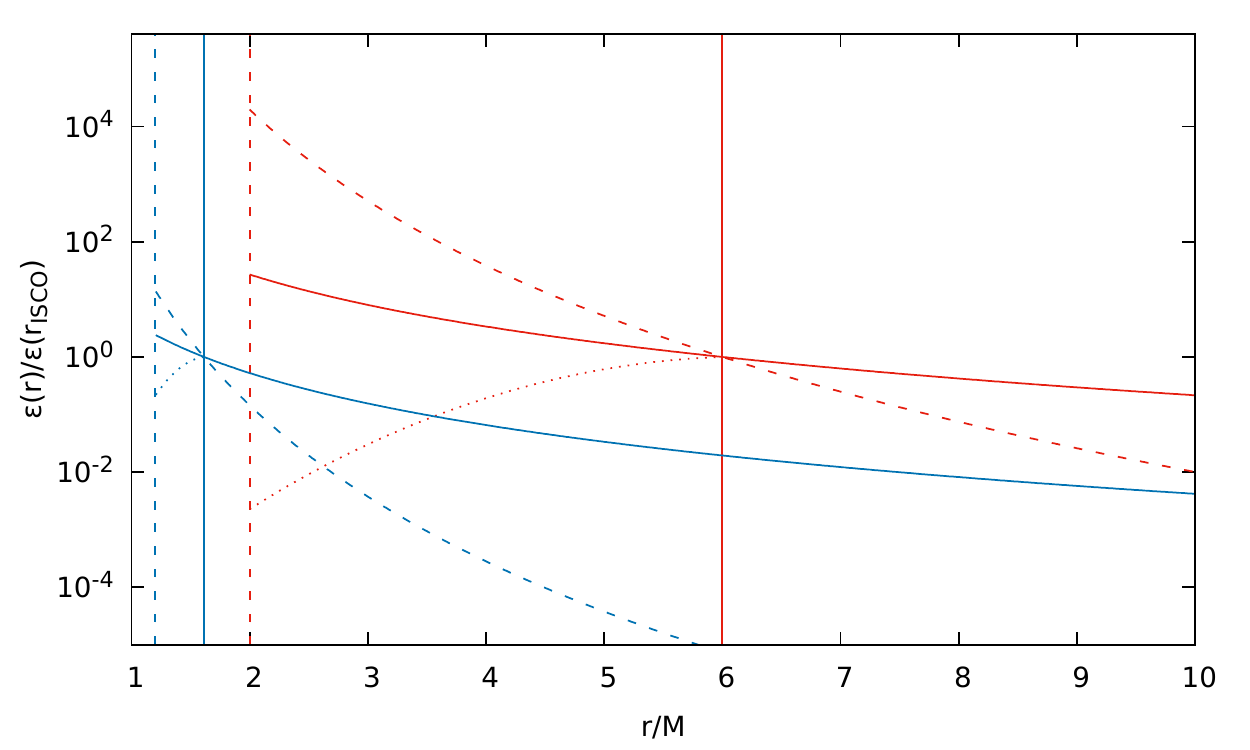}
\caption{Emissivity profile for a Kerr black hole with $\chi=0$ (red) and $\chi=0.98$ (blue). The solid curves are for an emissivity index $q=-3$ and the dashed curves for $q=-9$, Eq.~\ref{eq-emisd}. The dotted curves are the profile in the plunging region that falls off with smaller radius, Eq.~\ref{eq-emisp}. The emissivity profiles are scaled such that for a given spin they match at the ISCO radius. The vertical solid lines are the ISCO radius and the vertical dashed lines are the horizon radius. \label{fig-emissivity}}
\end{figure*}

Two different profiles will be used for the plunging region, the same power-law profile as above used in the thin disk region and a profile that falls off below the ISCO radius given by~\cite{C_rdenas_Avenda_o_2020}
\begin{equation}
\epsilon_{p} \propto \left(\frac{r}{r_{\text{ISCO}}}\right)e^{-\left(r-r_{\text{ISCO}}\right)^{2}/\left(r_{\text{ISCO}}-\ln r_{\text{ISCO}}-1\right)}. \label{eq-emisp}
\end{equation}
The emissivity profiles are shown in Fig.~\ref{fig-emissivity} for the Kerr spacetime with two values of dimensionless spin, $\chi=0$ and~$0.98$, and two values of the emissivity index, $q=-3$ and~$-9$.

The emissivity profile gives the relative intensity of emitted radiation from the disk, and since the absolute intensity is strongly dependent on the internal thermodynamical properties of the disk, here the intensity will simply be normalized by the maximum for any given configuration. By Liouville's theorem, $I_{\nu}/\nu^{3}$ is conserved along a photon's trajectory, where $\nu$ is the photon's frequency and $I_{\nu}$ is the specific intensity. The observed intensity is then related to the emitted intensity by the redshift
\begin{equation}
I_{\nu'}^{o}=g^{3}I_{\nu}^{e}.
\end{equation}
Integrating the specific intensity over frequency brings another factor of redshift and the total observed intensity is then
\begin{equation}
I^{o} \propto g^{4} \epsilon,
\end{equation}
where it has been used that the emissivity profile is proportional to the emitted intensity.

\section{Ray-tracing}
\label{sec:ray}

A general relativistic ray-tracing code is used to compute the shadow image and redshift factor associate with each photon trajectory. The ray-tracing code is based on the method described in~\cite{Psaltis_2011} and is a modified version of that used in~\cite{Ayzenberg_2018, Gott_2019}. As with massive particles in the previous section, all stationary and axisymmetric spacetimes have a conserved energy $E$ and angular momentum $L_{z}$ that can be related to components of the four-momentum, $p_{t}=-E$ and $p_{\phi}=L_{z}$. These relations lead to similar evolution equations as for massive particles (Eqs.~\ref{eq-dott} and~\ref{eq-dotphi}) for the $t$- and $\phi$-components of the photon position
\begin{align}
\frac{dt}{d\lambda'} = & -\frac{bg_{t\phi}+g_{\phi\phi}}{g_{tt}g_{\phi\phi}-g_{t\phi}^{2}}, \label{eq-dtdl}
\\
\frac{d\phi}{d\lambda'} = & \frac{g_{t\phi}+bg_{tt}}{g_{tt}g_{\phi\phi}-g_{t\phi}^{2}},
\end{align}
where $\lambda'\equiv E\lambda$ is the normalized affine parameter and $b\equiv L_{z}^{\gamma}/E^{\gamma}$ is, as before, a conserved quantity of the photon trajectory referred to as the impact parameter. Note that these are first-order differential equations owing to the symmetry of the spacetimes studied here. However, for the evolution of the $r$- and $\theta$-components the second-order geodesic equations are used such that the code is applicable to a wider range of spacetimes (although the Kerr and the Johannsen metrics have a Carter-like constant, are in turn separable, and the evolution equations can be written in purely first-order form for those spacetimes). These are
\begin{align}
\frac{d^{2}r}{d\lambda'^{2}} = & -\Gamma^{r}_{tt}\left(\frac{dt}{d\lambda'}\right)^{2}-\Gamma^{r}_{rr}\left(\frac{dr}{d\lambda'}\right)^{2}-\Gamma^{r}_{\theta\theta}\left(\frac{d\theta}{d\lambda'}\right)^{2}-\Gamma^{r}_{\phi\phi}\left(\frac{d\phi}{d\lambda'}\right)^{2}
\nonumber \\
& -2\Gamma^{r}_{t\phi}\left(\frac{dt}{d\lambda'}\right)\left(\frac{d\phi}{d\lambda'}\right)-2\Gamma^{r}_{r\theta}\left(\frac{dr}{d\lambda'}\right)\left(\frac{d\theta}{d\lambda'}\right),
\\
\frac{d^{2}\theta}{d\lambda'^{2}} = & -\Gamma^{\theta}_{tt}\left(\frac{dt}{d\lambda'}\right)^{2}-\Gamma^{\theta}_{rr}\left(\frac{dr}{d\lambda'}\right)^{2}-\Gamma^{\theta}_{\theta\theta}\left(\frac{d\theta}{d\lambda'}\right)^{2}-\Gamma^{\theta}_{\phi\phi}\left(\frac{d\phi}{d\lambda'}\right)^{2}
\nonumber \\
& -2\Gamma^{\theta}_{t\phi}\left(\frac{dt}{d\lambda'}\right)\left(\frac{d\phi}{d\lambda'}\right)-2\Gamma^{\theta}_{r\theta}\left(\frac{dr}{d\lambda'}\right)\left(\frac{d\theta}{d\lambda'}\right), \label{eq-d2thdl2}
\end{align}
where $\Gamma^{\alpha}_{\mu\nu}$ are the Christoffel symbols of the metric.

The coordinate system and reference frame are chosen such that the black hole is stationary at the origin and the black hole's spin angular momentum is along the $z$-axis. As the black hole mass only scales the shadow size and redshift value while keeping the shape and relative redshift of the image the same, units with the black hole mass set to unity, $M=1$, are used within the code and for the remainder of this paper. The observing screen is centered at a distance $D=10^{5}$, azimuthal angle $\theta=\iota$, and polar angle $\phi=0$, where $\iota$ is the inclination angle between the observing line-of-sight and the black hole's spin angular momentum ($z$-axis as chosen here). The polar coordinates $r_{\text{scr}}$ and $\phi_{\text{scr}}$ are used on the screen and can be related to the celestial coordinates $\left(\alpha,\beta\right)$ on the observer's sky by $\alpha=r_{\text{scr}}\cos\phi_{\text{scr}}$ and $\beta=r_{\text{scr}}\sin\phi_{\text{scr}}$.

The final positions and momenta of the photons impacting the observing screen are, in principle, known, and so the system of equations (Eqs.~\ref{eq-dtdl}-~\ref{eq-d2thdl2}) can be evolved backwards in time to find the locations of emission on the accretion disk. Each photon is initialized with some position on the screen and a four-momentum that is perpindicular to the screen, which simulates placing the screen at spatial infinity as only photons moving perpindicular to the screen at a finite distance will also hit a screen at spatial infinity.

The initial position and four-momentum of each photon is given by
\begin{align}
r_{i} = & \left(\alpha^{2}+\beta^{2}+D^{2}\right)^{1/2},
\\
\theta_{i} = & \arccos\left(\frac{D\cos\iota+\beta\sin\iota}{r_{i}}\right),
\\
\phi_{i} = & \arctan\left(\frac{\alpha}{D\sin\iota-\beta\cos\iota}\right),
\end{align}
and
\begin{align}
\left(\frac{dr}{d\lambda'}\right)_{i} = & \frac{D}{r_{i}},
\\
\left(\frac{d\theta}{d\lambda'}\right)_{i} = & \frac{-\cos\iota+\frac{D}{r_{i}^{2}}\left(D\cos\iota+\beta\sin\iota\right)}{\sqrt{r_{i}^{2}-\left(D\cos\iota+\beta\sin\iota\right)^{2}}},
\\
\left(\frac{d\phi}{d\lambda'}\right)_{i} = & \frac{-\alpha\sin\iota}{\alpha^{2}+\left(D\sin\iota-\beta\cos\iota\right)^{2}},
\\
\left(\frac{dt}{d\lambda'}\right)_{i} = & -\frac{g_{t\phi}}{g_{tt}}\left(\frac{d\phi}{d\lambda'}\right)_{i}
\nonumber \\
& +\sqrt{\frac{g_{t\phi}^{2}}{g_{tt}^{2}}\left(\frac{d\phi}{d\lambda'}\right)_{i}^{2}-\left[\frac{g_{rr}}{g_{tt}}\left(\frac{dr}{d\lambda'}\right)_{i}^{2}+\frac{g_{\theta\theta}}{g_{tt}}\left(\frac{d\theta}{d\lambda'}\right)_{i}^{2}+\frac{g_{\phi\phi}}{g_{tt}}\left(\frac{d\phi}{d\lambda'}\right)_{i}^{2}\right]},
\end{align}
where the final component, $\left(dt/d\lambda'\right)_{i}$, is found by requiring that the norm of the photon four-momentum is zero. The impact parameter $b$ is a conserved quantity and is computed from the initial conditions.

The system is solved using an adaptive 4th order Runge-Kutta method usually known as the Fehlberg method or RK45. Photons are evolved from the screen back to the black hole-disk system until one of three main conditions is met: 1) The photon crosses the event horizon (in reality the evolution is stopped a small distance outside the horizon in order to avoid the coordinate divergence at $r_{\text{H}}$), 2) the photon escapes to infinity,~i.e.~passes beyond the initial distance from the black hole, or 3) the photon crosses the equatorial plane for the 4th time (there are also various conditions to catch errors, numerical or otherwise). Each time the photon crosses the equatorial plane relevant physical values are calculated and recorded up to the 4th crossing,~i.e.~1.5 orbits of the black hole. The final information from each raytrace are the initial position on the screen, the radius on the disk at which the photon crossed, and the redshift of the photon had it been emitted from the location of the crossing.

The grid used on the screen for the photon initial positions is polar as the shadow image is roughly circular. The polar angle step size was chosen to roughly match two telescope resolutions at a distance of 3 gravitational radii from the center of the black hole. A step size of $\delta\phi_{\text{scr}}=2\pi/9$ is approximately a resolution of $10~\mu$as (on the order of the EHT resolution~\cite{2019ApJ...875L...1E}) and a smaller step size of $\delta\phi_{\text{scr}}=2\pi/90$ is approximately a resolution of $1~\mu$as (a factor of 10 improvement on the EHT, as may be possible with future space-based versions~\cite{Roelofs_2019}). As subsequent crossings, or subrings, are progressively smaller in radial extent, an adaptive grid is used for the radial coordinate on the screen. The radial grid step size is chosen to be $\delta r_{\text{scr}}=10^{-(N+1)}$, where $N$ is the number of crossings at a grid point. Thus, in a region on the screen where no photons are visible the step size is $\delta r_{\text{scr}}=0.1$ and in a region where there are the maximum of 4 crossings the step size is $\delta r_{\text{scr}}=10^{-5}$. This was found to be the optimal step size choice for computational time without sacrificing accuracy for the telescope resolutions studied. In order to match the telescope resolutions, the data is radially binned into bins of $\Delta r_{\text{scr}}=2$ for the $\sim 10~\mu$as resolution and $\Delta r_{\text{scr}}=0.2$ for the $\sim 1~\mu$as resolution.

\section{Analysis}
\label{sec:ana}

In order the determine how well black hole shadow observations can be used to estimate the physical parameters of a black hole and associated accretion disk, including non-Kerr modifications, a basic likelihood analysis is performed. The Kerr spacetime is assumed to be the correct description of black holes in nature,~i.e.~the Kerr shadow is the \textit{injected synthetic signal} or \textit{injection} for short. The Johannsen and KRZ shadows are the \textit{model} to be fit to the Kerr injection, noting that both the Johannsen and KRZ spacetimes include the Kerr spacetime when the deformation parameters vanish.

Due to computational limitations only one spin and inclination angle for the Kerr injection are analyzed, $\chi=0.94$ and $\iota=163^{\circ}$. These values, in particular the inclination angle, are in agreement (or at the least, not in disagreement) with the EHT observation of M87*~\cite{EventHorizonTelescope:2019pgp}. The mass of the injected black hole is set to 1, however other values are used for the model to allow the mass to take on a different value from the current measured value. Four different emissivity profiles are used for the injection, the two different profiles described in Sec.~\ref{sec:emprof} with emissivity indices $q=-3$ and $q=-9$. These are used to represent the uncertainty in the correct accretion disk model, although realistic accretion disk models are much more complicated. So to summarize, four injections are used where the only difference between the injections is the emissivity profile. While this is a limited set for the Kerr injection, other values, particularly lower and higher spins and lower values of inclination angle, were briefly studied and the results were found to be qualitatively similar regardless of spin or inclination angle.

The range of values used for the model are centered around the Kerr injection values as the deviation from Kerr is not expected to be large. The ranges are smaller for the $\sim 1~\mu$as resolution case as producing these shadow images is more computationally expensive and with a higher resolution the constraints should be stronger. For the $\sim 1~\mu$as resolution, shadow images are analyzed for dimensionless spins $\chi=[0.91,0.97]$ in steps of $\delta\chi=0.1$, inclination angles $\iota=[160^{\circ},166^{\circ}]$ in steps of $\delta\iota=1^{\circ}$, masses $M=[0.9,1.1]$ in steps of $\delta M=0.01$ (where the injected mass is $M=1$), emissivity indices $q=[-9,-1]$ in steps of $\delta q=1$ (for both described emissivity profiles), and scaled deformation parameters $\mu=[-0.3,0.3]$ in steps of $\delta\mu=0.1$\footnote{The scaled deformation parameter is defined such that $\mu=\pm1$ is either equal to $\pm1$ of the relevant deformation parameter or the maximum/minimum allowed value (as given by Eqs.~\ref{eq-a13bound}-\ref{eq-a22bound}+\ref{eq-d1bound}-\ref{eq-d6bound}) $\mp0.2$, whichever is closer to $0$. The shift of $\mp0.2$ is done to avoid any odd behavior that may occur near the border regions of the allowed values.}. For the $\sim 10~\mu$as resolution the ranges for the masses and emissivity indices are the same and the other ranges are dimensionless spins $\chi=[0.84,0.99]$ in steps of $\delta\chi=0.1$, inclination angles $\iota=[153^{\circ},173^{\circ}]$ in steps of $\delta\iota=1^{\circ}$, and scaled deformation parameters $\mu=[-1.0,1.0]$ in steps of $\delta\mu=0.1$. A total of 4,900 ray-tracing simulations at the $\sim 1~\mu$as resolution and 70,560 ray-tracing simulations at the $\sim 10~\mu$as resolution are performed, only dependent on the spin, deformation parameter, and inclination angle. The emissivity profiles are incorporated during the binning process as they do not impact the ray-tracing and the masses are incorporated during the analysis as they only scale the image. Overall, 1,852,200 shadow images at the $\sim 1~\mu$as resolution and 26,671,680 shadow images at the $\sim 10~\mu$as resolution are analyzed.

For the analysis, three different realizations of the data are studied in order to determine if focusing on different aspects of the shadow image may be better for extracting parameters of the spacetime and/or accretion disk. Studying the different realizations is also useful since a given realization may not be observable in reality due to telescope limitations or more complex accretion disk physics that is not included here. The three realizations are: 1) the full image with and without the $n=0$ subring, 2) the screen angle dependent radius of each subring, where the brightest point of the subring for a given screen angle is chosen as the radius, and 3) the ratio of the screen angle dependent radius between sequential subrings. 

\begin{figure*}[hpt]
\includegraphics[width=0.5\columnwidth{},clip=true]{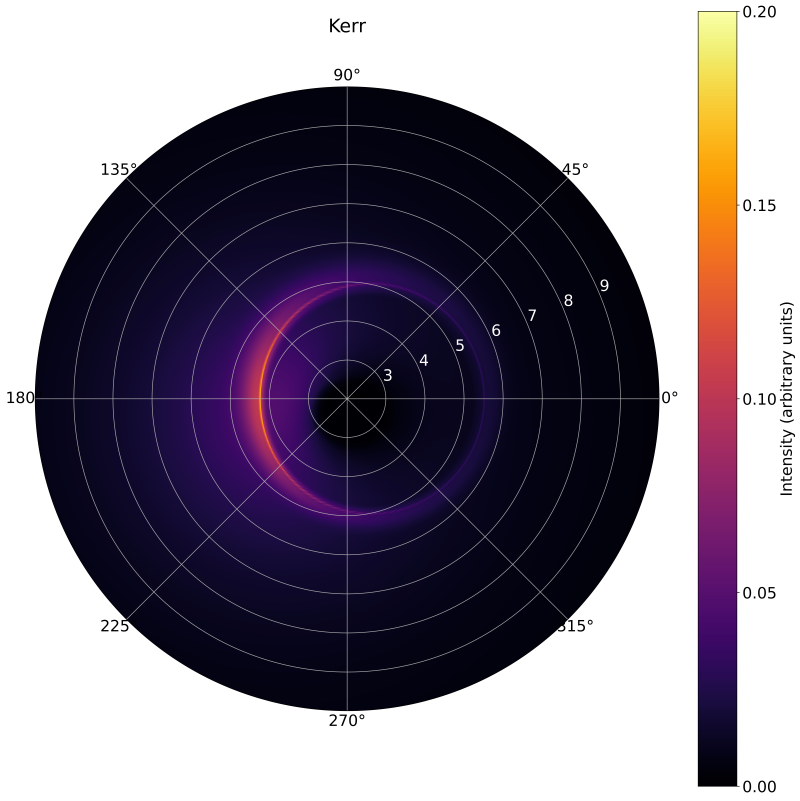}
\includegraphics[width=0.5\columnwidth{},clip=true]{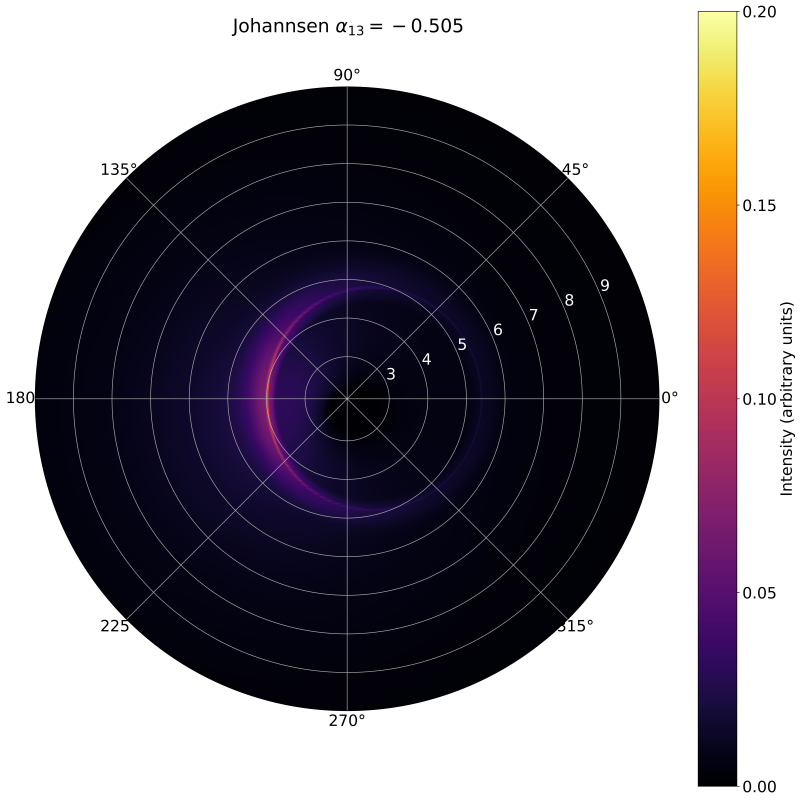}
\\
\includegraphics[width=0.5\columnwidth{},clip=true]{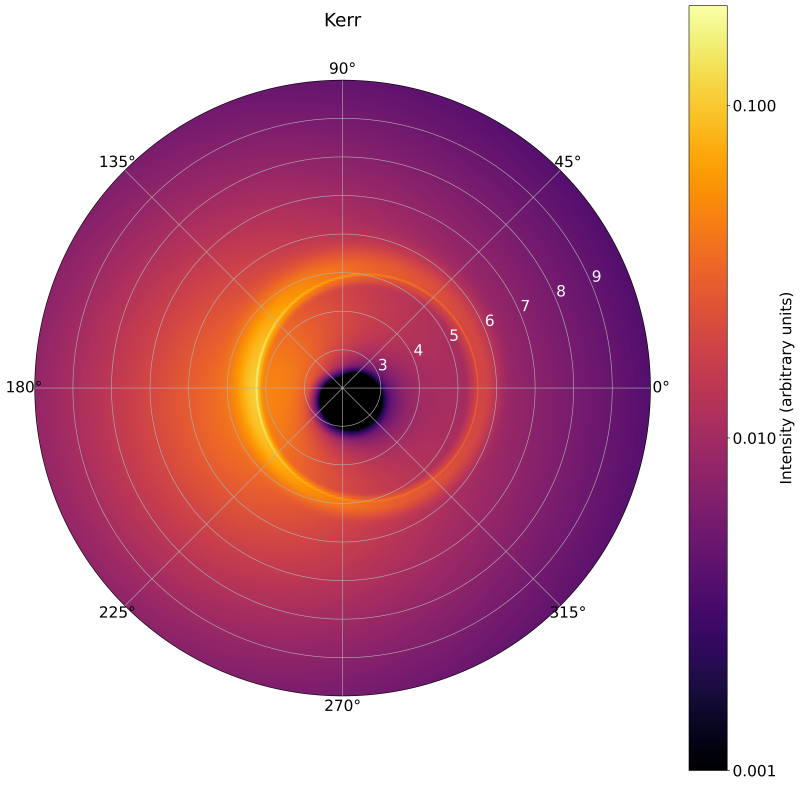}
\includegraphics[width=0.5\columnwidth{},clip=true]{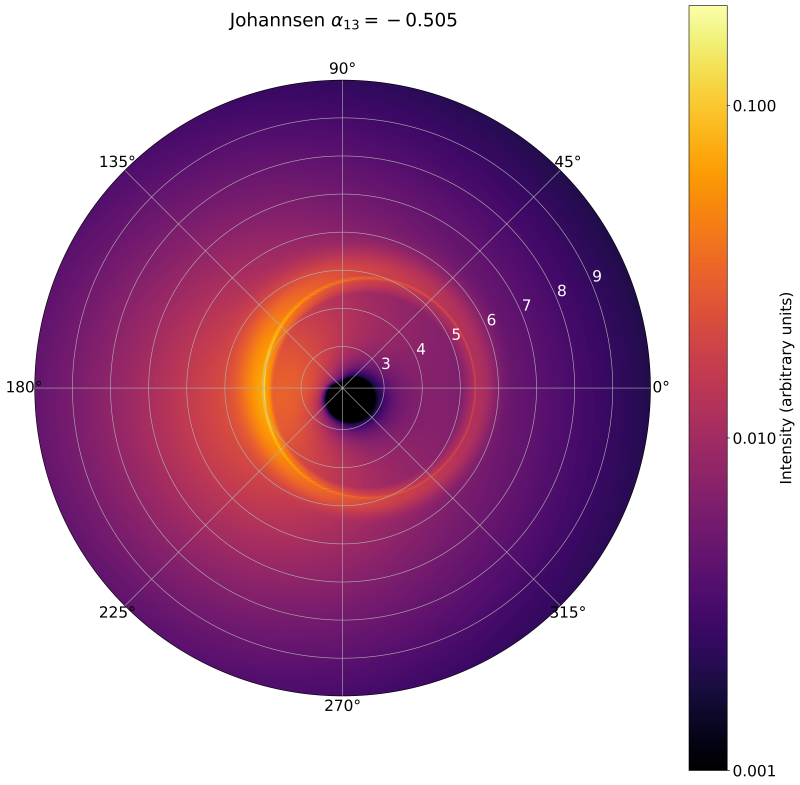}
\caption{Shadow images (top has intensity on a linear scale, bottom on a logarithmic scale) for Kerr (left) and Johannsen (right) black holes with dimensionless spin $\chi=0.94$, inclination angle $\iota=163^{\circ}$, and using the emissivity profile that decreases below the ISCO with emissivity index $\epsilon=-3$. The Johannsen black hole is with a non-zero $\alpha_{13}=-0.505$ deformation parameter, which is the largest negative value used in this work for $\chi=0.94$. \label{fig:shad}}
\end{figure*}

\begin{figure*}[hpt]
\includegraphics[width=0.5\columnwidth{},clip=true]{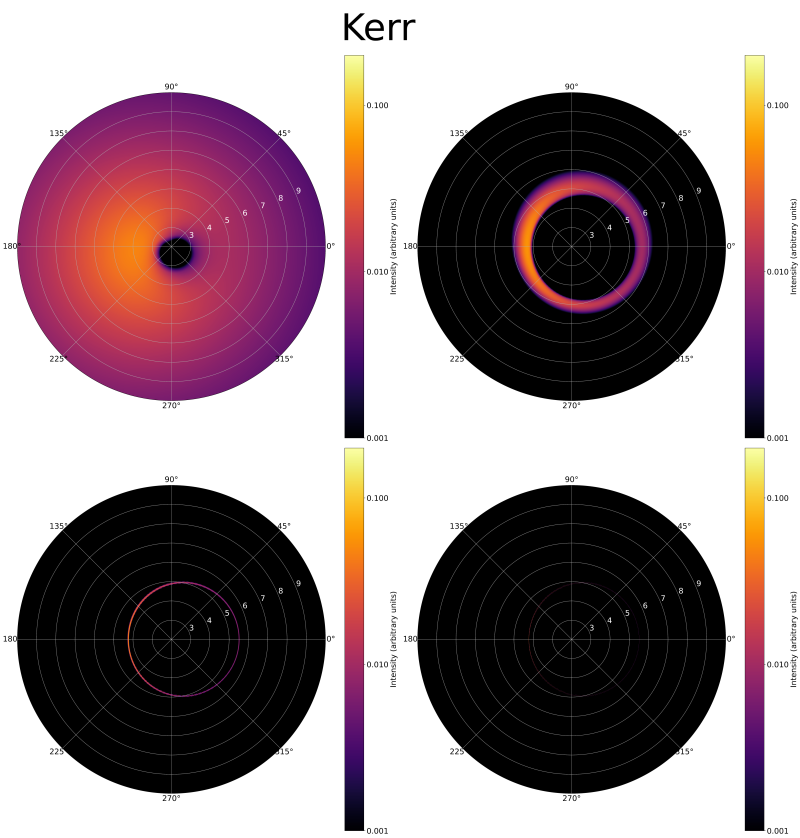}
\includegraphics[width=0.5\columnwidth{},clip=true]{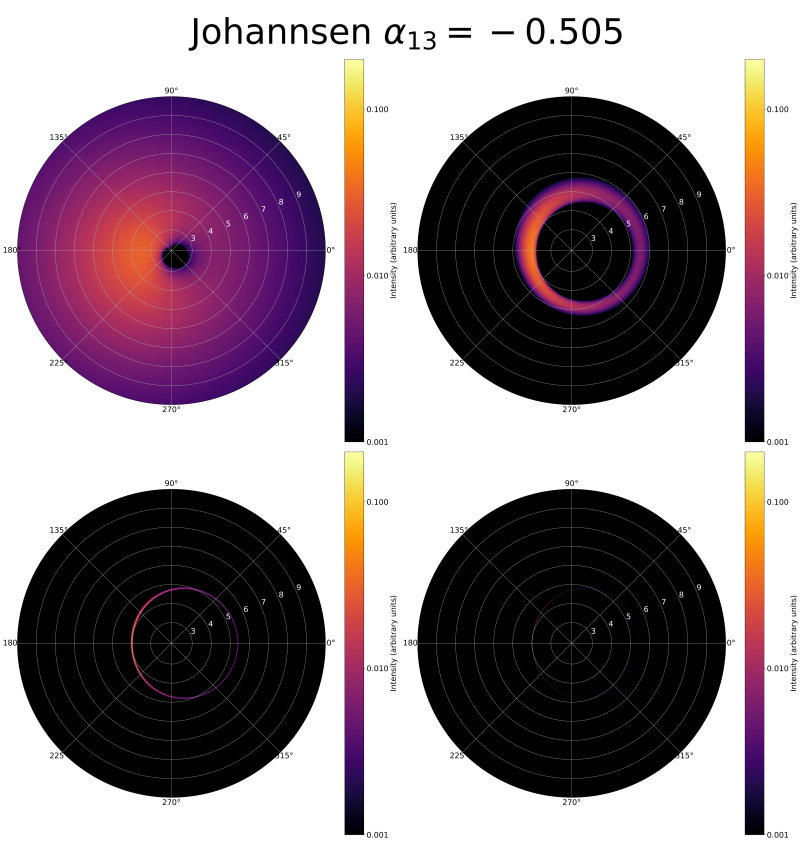}
\\
\includegraphics[width=0.5\columnwidth{},clip=true]{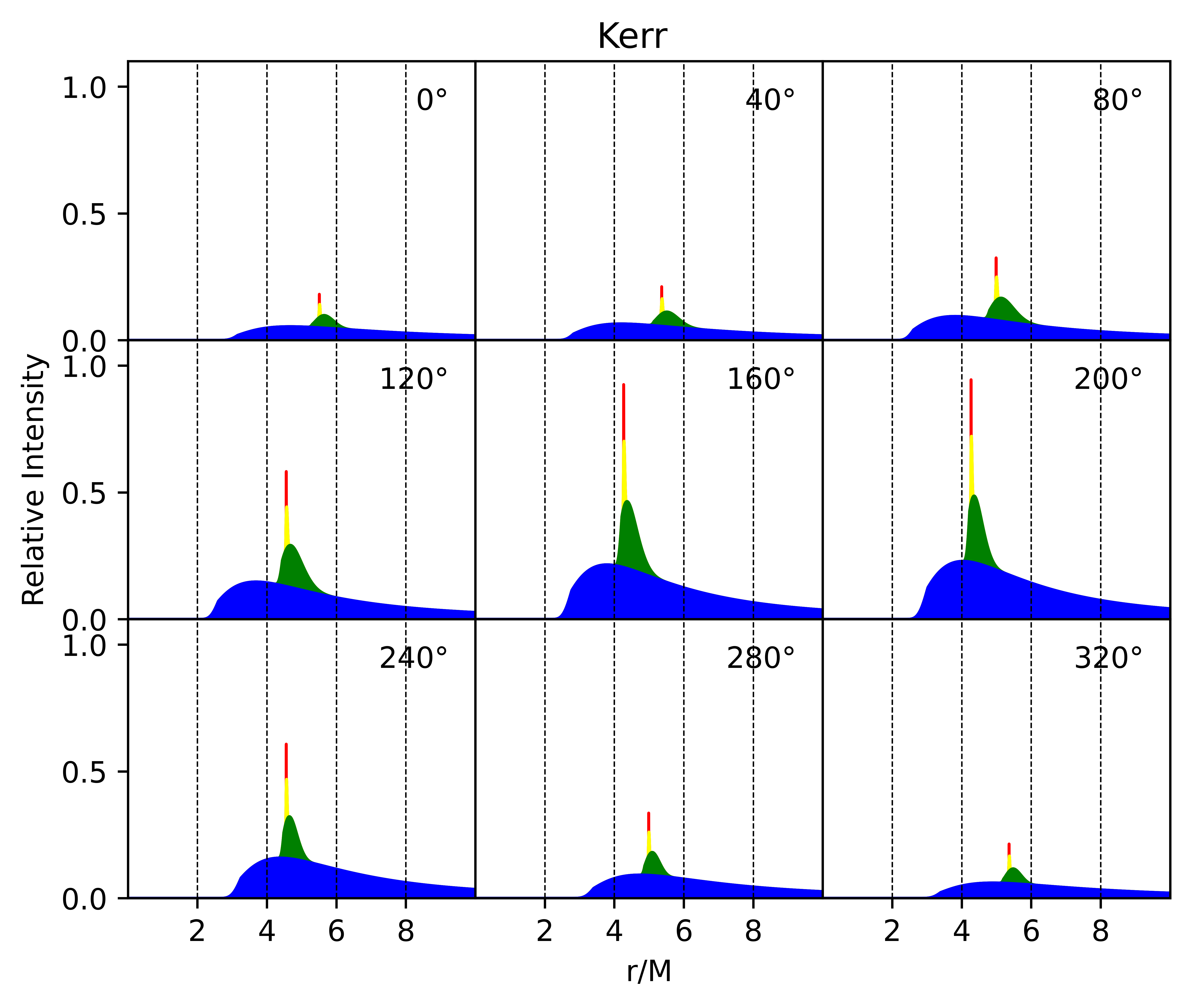}
\includegraphics[width=0.5\columnwidth{},clip=true]{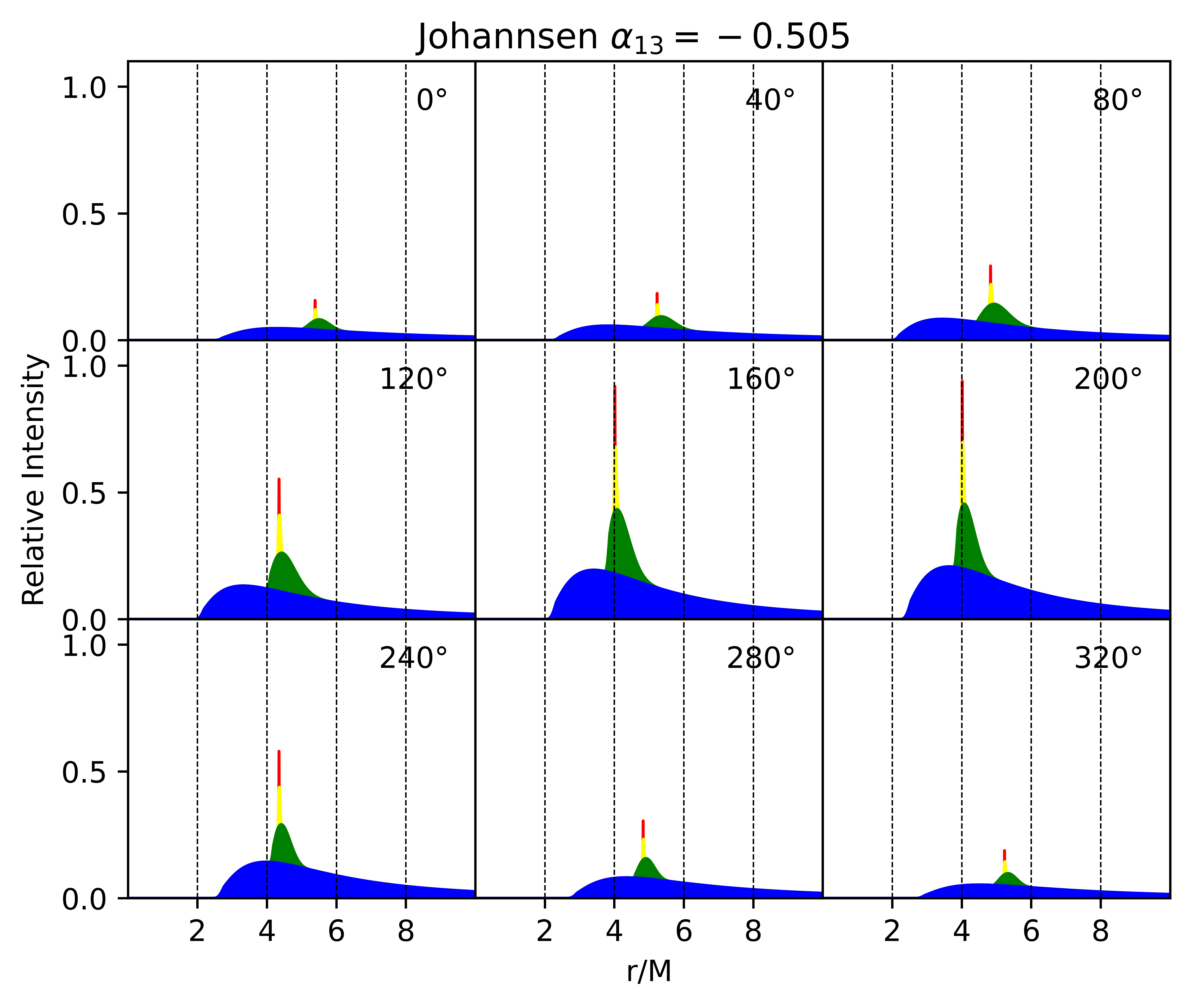}
\caption{Top: Subrings of the shadow image with intensity on a log scale for Kerr (left) and Johannsen (right) black holes with dimensionless spin $\chi=0.94$, inclination angle $\iota=163^{\circ}$, and using the emissivity profile that decreases below the ISCO with emissivity index $\epsilon=-3$. The Johannsen black hole is with a non-zero $\alpha_{13}=-0.505$ deformation parameter, which is the largest negative value used in this work for $\chi=0.94$. Top left is the n=0 subring, top right n=1, bottom left n=2, and bottom right n=3. \\ Bottom: Intensity of the shadow subrings along different screen angles for Kerr (left) and Johannsen (right). Intensity is scaled to the maximum in the image. The blue shaded region is the contribution from the n=0 subring, green n=1, yellow n=2, and red n=3. Note the intensities are stacked and only the visible shaded region represents the intensity contributed by each subring. \label{fig:split}}
\end{figure*}

\begin{figure*}[hpt]
\includegraphics[width=0.5\columnwidth{},clip=true]{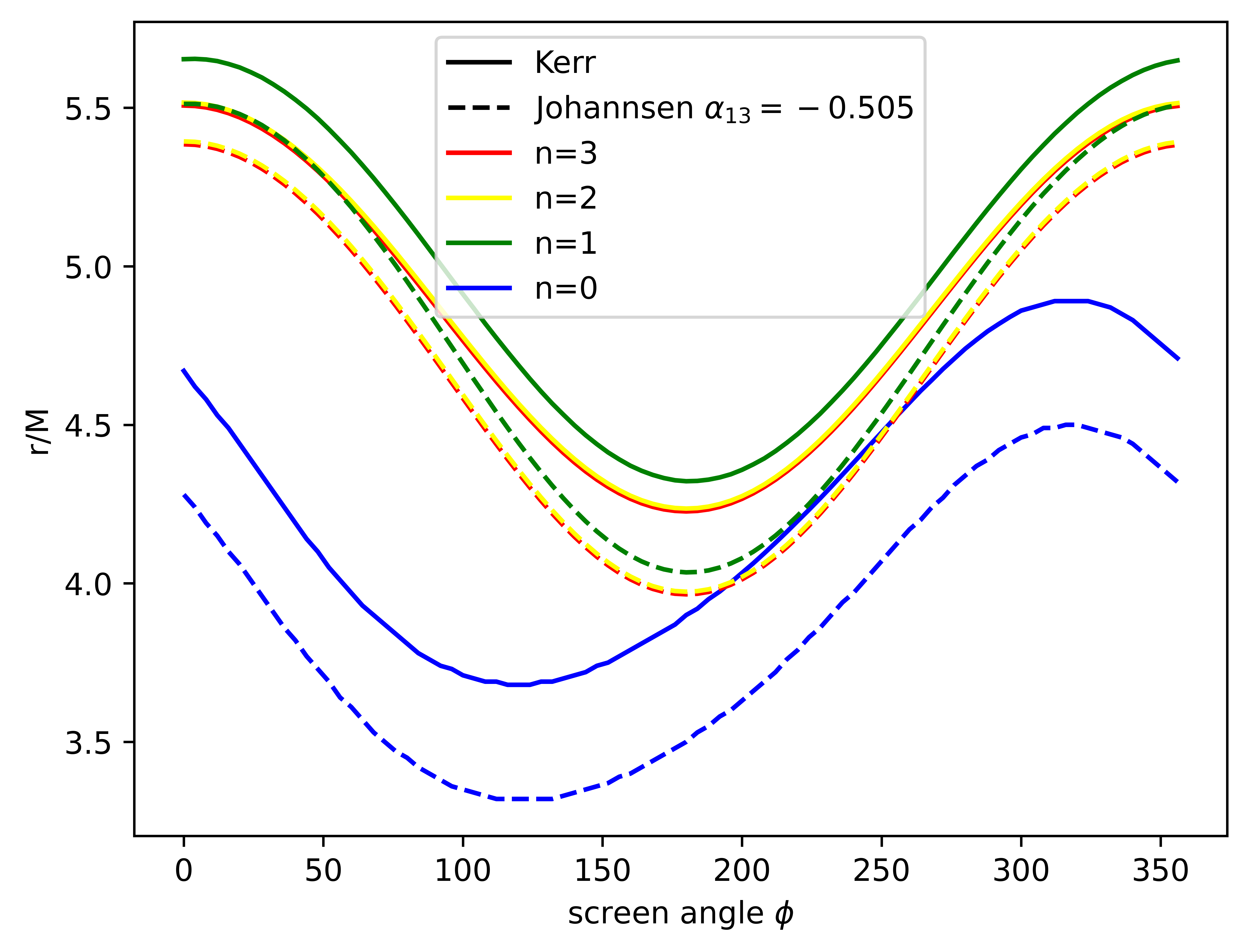}
\includegraphics[width=0.5\columnwidth{},clip=true]{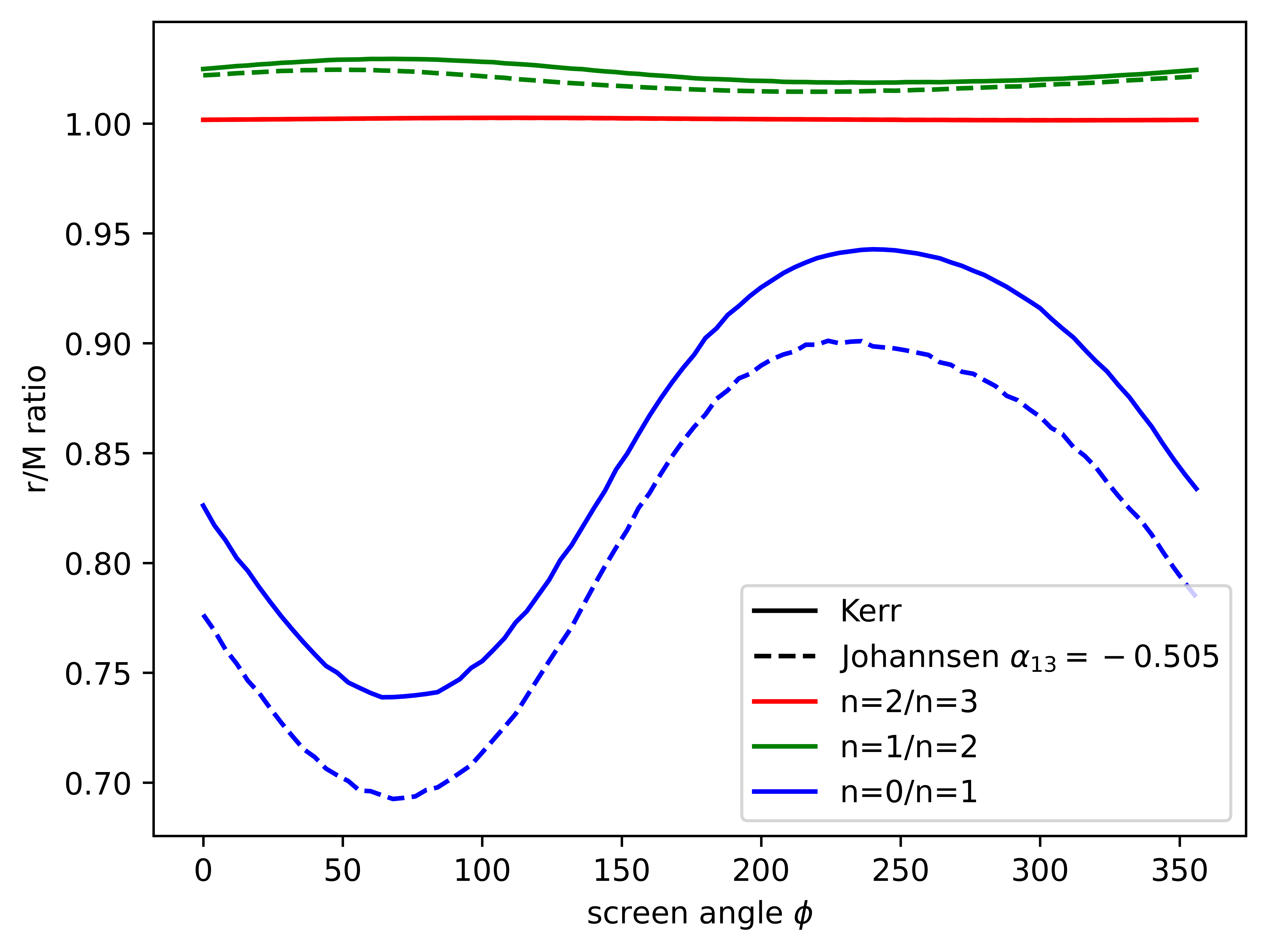}
\caption{Left: Radius of the brightest point in each subring as a function of screen angle for Kerr (solid) and Johannsen (dashed) black holes with dimensionless spin $\chi=0.94$, inclination angle $\iota=163^{\circ}$, and using the emissivity profile that decreases below the ISCO with emissivity index $\epsilon=-3$. The Johannsen black hole is with a non-zero $\alpha_{13}=-0.505$ deformation parameter, which is the largest negative value used in this work for $\chi=0.94$. Right: Ratio of the radius of the brightest point in each subring of consecutive subrings. \label{fig:ring}}
\end{figure*}

Figure~\ref{fig:shad} shows a comparison of the shadow images for Kerr and Johannsen with a non-zero $\alpha_{13}$ deformation parameter. Figure~\ref{fig:split} shows a comparison of the image split up into subrings as well as the relative luminosity of the subrings along different screen angles. Note that in these figures the resolution is quite high and there is some smoothing in generating the figures, while in the analysis the data is binned into relatively large spatial bins as described in the previous section. Figure~\ref{fig:ring} shows a comparison of the radius of the brightest point of the subrings and the ratio of the radii of sequential subrings. The differences between the Kerr and Johannsen shadow images are fairly minor, even though the value for $\alpha_{13}$ used for the Johannsen image is the largest negative value used in this work for the dimensionless spin of $\chi=0.94$,~i.e.~$\alpha_{13}=-0.505$, which is 0.2 above the limit set by Eq.~\ref{eq-a13bound}. The radius of all the subrings is slightly smaller and the intensity is slightly reduced in the Johannsen case, but it is clear that without very high resolution images, as in the figures, it would be quite difficult to distinguish between these two cases. Note Fig.~\ref{fig:ring} shows expected behavior as the higher order rings approach a value that is set by the location of the ideal photon sphere,~i.e.~the $n=\infty$ subring. This is even more clear in the plot of the ring radius ratios as the ratio of $n=2/n=3$ is about unity and this is the case for both the Kerr and Johannsen black holes.


A likelihood analysis is performed to calculate the posterior probability distributions of the model parameters for the four injections, where the injections only differ in the emissivity profile. Both the 1-dimensional and 2-dimensional posterior probabilities are calculated,~i.e.~marginalizing over all but 1 and 2 parameters, respectively. As an example, the 1D posterior probability for spin is explicitly
\begin{equation}
P(\chi)=\int \mathcal{L}(\chi,\iota,\epsilon,M,\zeta)p(\iota)p(q)p(M)p(\zeta)d\iota dq dM d\zeta,
\end{equation}
where $\mathcal{L}$ is the likelihood function given by
\begin{equation}
\mathcal{L}(\chi,\iota,q,M,\zeta)\propto e^{-\chi^{2}(\chi,\iota,q,M,\zeta)/2},
\end{equation}
with the chi-squared $\chi^{2}$ and prior probability distributions $p$. The parameters used here are the dimensionless spin $\chi$, the inclination angle $\iota$, the relative mass $M$ (where the injection has $M=1$), the emissivity index $q$ (including the emissivity profile), and the deformation parameter here denoted as $\zeta$. For all parameters a flat prior probability distribution is used. The chi-squared is given by
\begin{equation}
\chi^{2}=\sum_{i}\left(\frac{o^{i}_{\text{mod}}-o^{i}_{\text{inj}}}{\sigma_{i}}\right)^{2},
\end{equation}
where the sum is over all data elements (bins in the case of the full shadow image or screen angle in the case of the subring radius or radius ratio), $o^{i}$ is the observable for each data element (intensity in the case of the full shadow image, radius for the subring radius, or radius ratio for the subring radius ratio), the subscripts ``mod" and ``inj" represent the model and injection, respectively, and $\sigma$ is the error in the observable for the given data element. The error for the full shadow image is calculated by averaging the difference in intensity between the current bin and the four adjacent bins. This is chosen for simplicity as a representation of the error due to having a limited resolution; calculating the actual observational error would be significantly more complicated. The error for the subring radius and subring radius ratio is simply the chosen bin resolutions, so $2$ or $0.2$ for the two resolutions used.

\section{Results}
\label{sec:results}

In general, the results of the analysis are qualitatively similar regardless of the non-Kerr metric modeled, so the results for the Johannsen metric with a non-zero $\alpha_{13}$ parameter will be specifically presented and discussed with additional comments about any qualitative departures from these results for the KRZ metric and other parameters.

\subsection{Full shadow image}

For the full shadow image with a resolution of $\sim 1~\mu$as, which is a resolution that may be possible with a future space-based version of the EHT, the results are quite promising. When the $n=0$ subring is included the injected parameters are exactly recovered, within the step-size as presented at the start of this section, and with a posterior probability that is very tightly peaked at the injection. The disk physics is certainly much more complex than simulated here and the uncertainty in the correct disk model may not be properly addressed here. In an attempt to account for this, the analysis is repeated with the $n=0$ subring removed. This subring is the most influenced by the disk physics, as it is the direct image of the disk, and removing it should reduce the influence of the disk physics on the extraction of the spacetime parameters. Note that separating the $n=0$ subring from the actual observation is, in principle, possible (see~e.g.~\cite{Tiede:2020iif, Gelles:2021oyd}). With the $n=0$ subring excluded, the parameters are not recovered as well, but still better than what is possible with other current observations. In general, the mass and inclination angle are recovered exactly, as before. When the emissivity index is $q=-3$, the analysis cannot distinguish between the two emissivity models, but does so when $q=-9$, likely because with the less steep emissivity profile the intensity is not so strongly concentrated around the ISCO. The spin is recovered to within two steps,~i.e.~$\pm0.2$, while the deformation parameter is also usually recovered to within $\pm0.2$ (where here the scaled deformation parameter is used). For two cases, Johannsen with non-zero $\alpha_{52}$ and non-zero $\epsilon_{3}$, the posteriors peak around zero (the Kerr value), but it would not be possible to place constraints with the range of deformation parameters used for this high resolution case. This is not unexpected as the effect on the spacetime due to $\alpha_{52}$ and $\epsilon_{3}$ is weaker than from other deformation parameters (e.g.~$\alpha_{52}$ doesn't modify the ISCO radius and $\epsilon_{3}$ only modifies the ISCO radius slightly as in Fig.~\ref{fig:iscoJo}). Overall, the results at the higher resolution of $\sim 1~\mu$as suggest that a space-based version of the EHT would be able to place very strong constraints on departures from the Kerr metric.

\begin{figure*}[hpt]
\includegraphics[width=0.5\columnwidth{},clip=true]{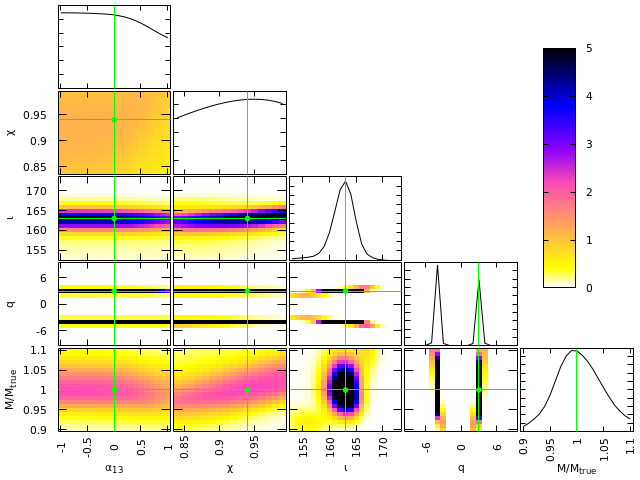}
\includegraphics[width=0.5\columnwidth{},clip=true]{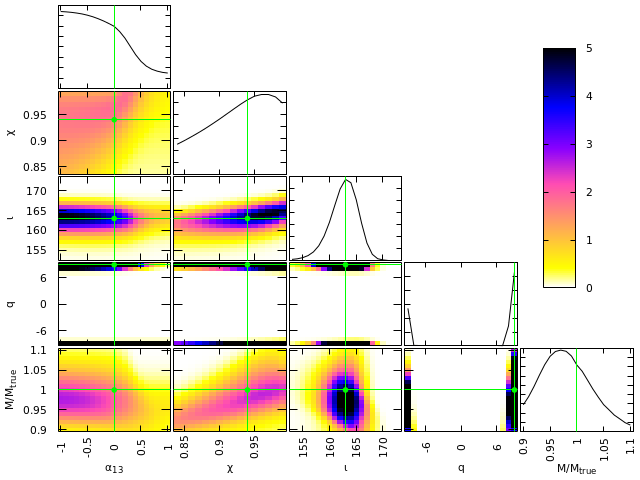}
\caption{Posterior probabilities for the full image with a $\sim10~\mu$as resolution for a Kerr injection with an emissivity profile that falls off below the ISCO with emissivity indices $q=-3$ (left) and $q=-9$ (right), recovered with a Johannsen model with a non-zero $\alpha_{13}$ deformation parameter. In the plots positive values of the emissivity index $q$ are for the emissivity profile that falls off below the ISCO, but the actual value used is negative. The green lines and points mark the injected parameter values. \label{fig:Joa13_10}}
\end{figure*}

The results are less positive in the case of the full shadow image with a resolution of $\sim 10~\mu$as. Figure~\ref{fig:Joa13_10} shows the posterior probabilities for the full image for a Kerr injection with an emissivity profile that falls off below the ISCO radius for emissivity indices $q=-3$ and $q=-9$ when recovered with the Johannsen metric with a non-zero $\alpha_{13}$ deformation parameter. The results are qualitatively similar for the two different emissivity profiles and for all deformation parameters except $\alpha_{52}$ and $\epsilon_{3}$ of the Johannsen metric. As in the high resolution case, $\alpha_{52}$ and $\epsilon_{3}$ have a weaker effect on the image and the resulting 1D posterior probabilities for the deformation parameters remain close to the flat prior, so they cannot be constrained at all. For the other deformation parameters in both metrics the posteriors tend to favor negative or positive values, as seen in Fig.~\ref{fig:Joa13_10} or peak near the Kerr value of $0$ and fall off towards both sides. Similarly, for the posteriors of the dimensionless spin $\chi$, the probability tends to peak at or near the injected value and fall off to either side. However, it is worth noting that in some cases (both for different metrics/deformation parameters and emissivity profiles/indices) the value at the peak of the posterior for the deformation parameter and spin is not the injected value. This along with the 2D posterior probabilities suggest there is fairly strong degeneracy between the spin and deformation parameter in some cases, which is to be expected. Both the spin and deformation parameter are parameters that are only apparent very near the black hole and can have similar influence on the spacetime. The emissivity index is always well-recovered, but the analysis is not able to decisively determine which emissivity profile is used. Both the inclination angle and mass are usually well-recovered, but in some cases the most likely value is slightly shifted from the injection. Again, this is likely due to some degeneracies between all of the parameters, though these degeneracies are weaker than that between the spin and deformation parameter.

\begin{figure*}[hpt]
\includegraphics[width=0.5\columnwidth{},clip=true]{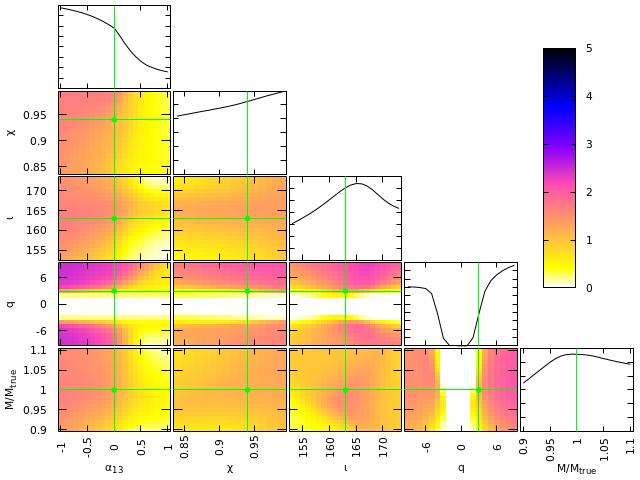}
\includegraphics[width=0.5\columnwidth{},clip=true]{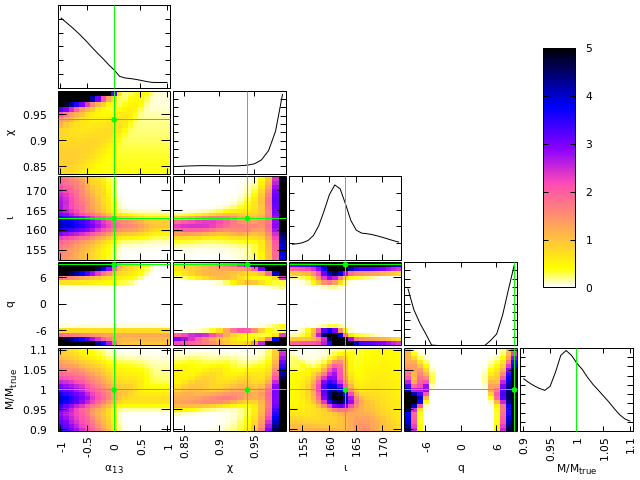}
\caption{Posterior probabilities for the full image with a $\sim~10~\mu$as resolution with the $n=0$ subring removed for a Kerr injection with an emissivity profile that falls off below the ISCO with emissivity indices $q=-3$ (left) and $q=-9$ (right), recovered with a Johannsen model with a non-zero $\alpha_{13}$ deformation parameter. In the plots positive values of the emissivity index $q$ are for the emissivity profile that falls off below the ISCO, but the actual value used is negative. The green lines and points mark the injected parameter values. \label{fig:Joa13_10_0}}
\end{figure*}

Results for the lower resolution full image with the $n=0$ subring removed are shown in Fig.~\ref{fig:Joa13_10_0}. With the $n=0$ subring the results are dependent on the emissivity index $q$. When the emissivity index is $q=-3$, all posteriors become flatter than when the $n=0$ subring is included. Additionally the fits favor a higher emissivity index. When the emissivity index is $q=-9$ the dimensionless spin and deformation parameter pin to values that have a smaller ISCO radius (so high spin $\chi\approx0.99$ and either high or low deformation parameter $\zeta\pm1$ when the deformation parameter modifies the ISCO radius). In these cases the inclination angle, emissivity index/profile, and mass are qualitatively similar to when the $n=0$ subring is included. It's unclear why this pinning towards smaller ISCO values is occurring. Perhaps images with more of the intensity focused towards smaller radii are preferred, which would happen with a lower ISCO, and there is a degeneracy that allows this without significantly modifying the image from the injection otherwise.

Overall, the full shadow image with the lower resolution of $\sim 10~\mu$as suggests that even with the current EHT and future earth-based versions of the EHT, some constraints can be placed on the spin and departures from the Kerr solution, though the spin constraints will be fairly weak compared to other current methods and the constraints on non-Kerr spacetimes will be roughly comparable. These results do point out that one must be careful as degeneracies between the physical parameters can lead to incorrect estimates even of the mass. In particular, when the $n=0$ subring is removed estimates of the spin can be far from the injected value.

\subsection{Subring radius}

As can be expected, analyzing only the radius of the subrings results in weaker constraints than the full image. However, it is quite possible that the accretion disk model used here is much too simple and in reality the disk cannot be modeled well enough to use the full image as a reliable observable for extracting the parameters of the spacetime. In such a case, one is left with analyzing the subrings, which quickly become effectively independent of the accretion disk physics as the order of the subring increases. Unfortunately, in the case of the lower resolution of $\sim10~\mu$as no constraints can be placed on any of the parameters except the emissivity index where there are weak constraints for a high emissivity index $q=-9$ and fairly strong constraints for a low emissivity index $q=-3$. These results are not surprising as the error in the subring radius at this resolution is quite high, $\sigma_{r}=2$. Now the analysis conducted here is fairly simple and doesn't take into account how the very long baseline interferometry of the EHT actually functions and how the data is processed. In reality, it may be possible to achieve a lower error in the subring radius than the effective resolution of the interferometer (see~e.g.~\cite{Tiede:2020iif, Gelles:2021oyd, Younsi:2021dxe}). If that is so, one can consider the error in the radius lower and even optimistically that it will approach the $\sigma_{r}=0.2$ of the $\sim 1~\mu$as resolution.

\begin{figure*}[hpt]
\includegraphics[width=0.5\columnwidth{},clip=true]{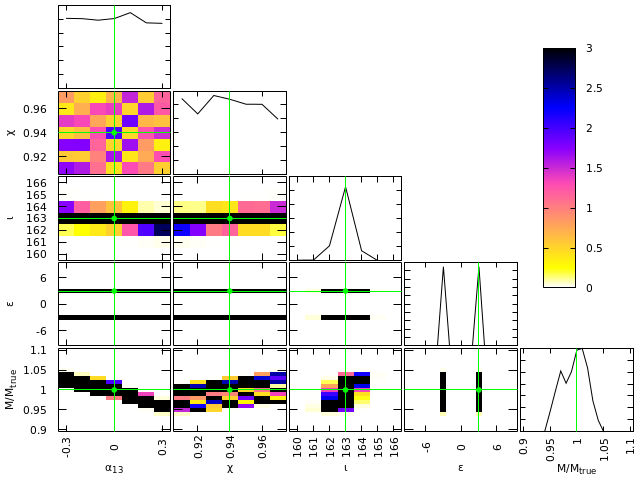}
\includegraphics[width=0.5\columnwidth{},clip=true]{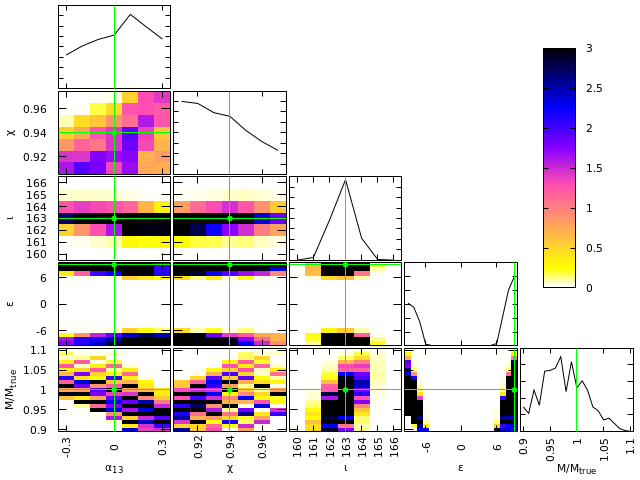}
\\
\includegraphics[width=0.5\columnwidth{},clip=true]{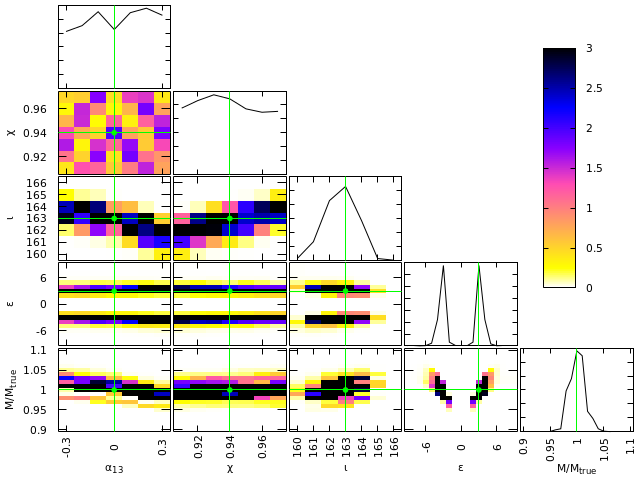}
\includegraphics[width=0.5\columnwidth{},clip=true]{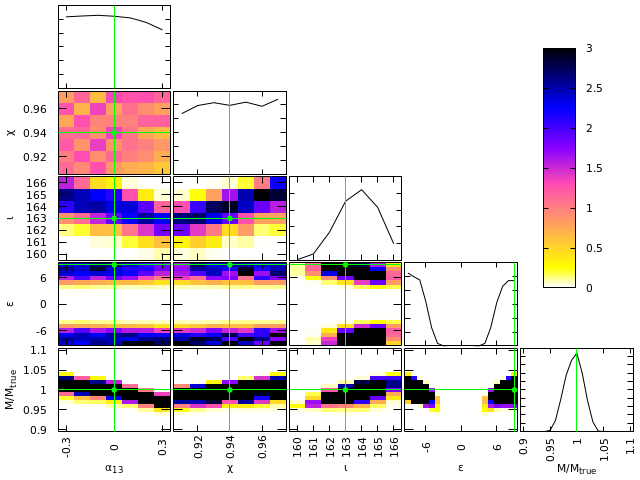}
\caption{Posterior probabilities for the subring radius observable (top is the $n=0$ subring and bottom is the $n=1$ subring) with a $\sim 1~\mu$as resolution for a Kerr injection with an emissivity profile that falls off below the ISCO with emissivity indices $q=-3$ (left) and $q=-9$ (right), recovered with a Johannsen model with a non-zero $\alpha_{13}$ deformation parameter. In the plots positive values of the emissivity index $q$ are for the emissivity profile that falls off below the ISCO, but the actual value used is negative. The green lines and points mark the injected parameter values. \label{fig:Joa13ring}}
\end{figure*}

Figure~\ref{fig:Joa13ring} shows the posterior probabilities when analyzing the subring radius for both the $n=0$ and $n=1$ subrings, again when the model is the Johannsen spacetime with a non-zero $\alpha_{13}$ deformation parameter. The results are qualitatively consistent across deformation parameters and spacetimes. In all cases constraints on the spin and deformation parameter are fairly weak or nonexistent, keeping in mind that the range of spins and deformation parameters modeled in this high resolution case is small. Previous work (e.g.~\cite{Broderick:2021ohx}) suggested the spin could be measured fairly accurately with a subring measurement, however the precision in the subring measurement in those works was at least an order of magnitude higher than used in this work. Beyond these two spacetime parameters, the $n=1$ subring in particular does a fairly good job of recovering and constraining the inclination angle, emissivity index, and mass. The $n=0$ subring does not recover or constrain the mass very well, likely due to the influence of the disk physics, so one must be careful when claiming anything about the mass with this observable. The specific emissivity profile,~i.e.~whether it is the profile that falls off below the ISCO or not, is not well constrained, but that is to be expected as the just the subring radius carries little information about the overall disk physics. The subring is certainly useful as an observable, however the results presented here and work from others suggest it is not particularly useful for constraining strong-gravity spacetime parameters,~i.e.~the spin and deformation from Kerr, unless the precision in the measurement is very high.

%

\subsection{Subring radius ratio}

\begin{figure*}[hpt]
\includegraphics[width=0.5\columnwidth{},clip=true]{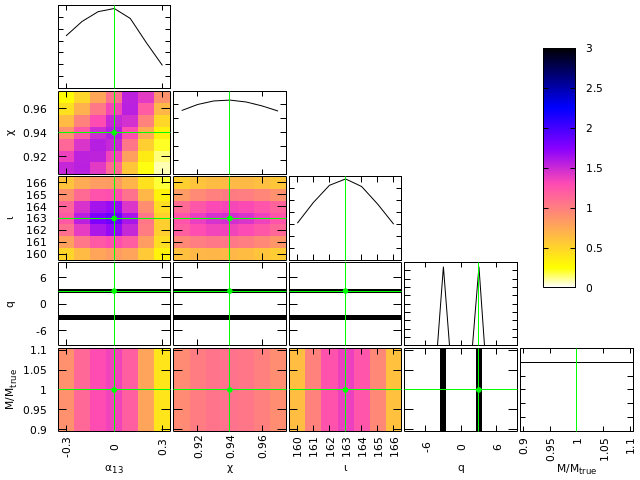}
\includegraphics[width=0.5\columnwidth{},clip=true]{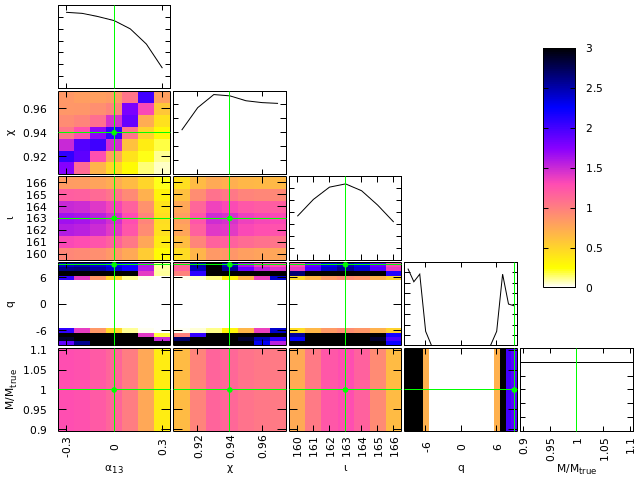}
\\
\includegraphics[width=0.5\columnwidth{},clip=true]{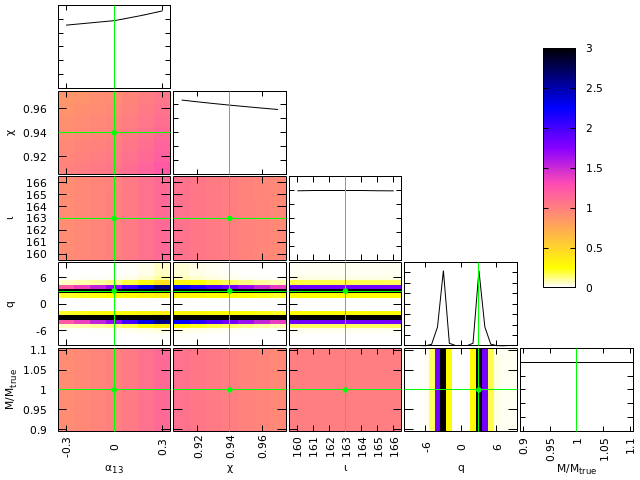}
\includegraphics[width=0.5\columnwidth{},clip=true]{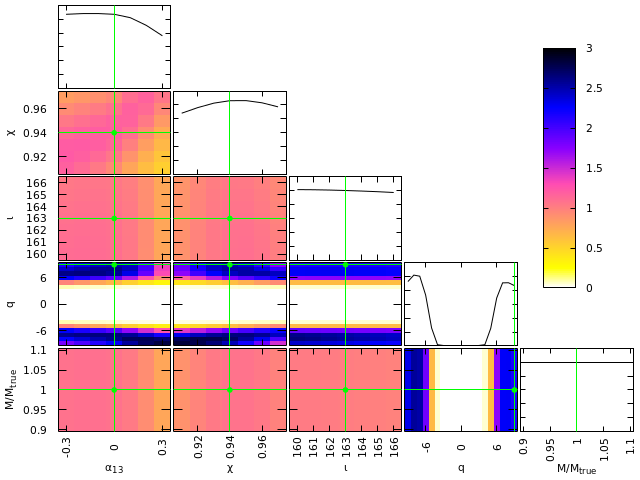}
\caption{Posterior probabilities for the subring radius ratio observable (top is the $n=0/n=1$ subring ratio and bottom is the $n=1/n=2$ subring ratio) with a $\sim 1~\mu$as resolution for a Kerr injection with an emissivity profile that falls off below the ISCO with emissivity indices $q=-3$ (left) and $q=-9$ (right), recovered with a Johannsen model with a non-zero $\alpha_{13}$ deformation parameter. In the plots positive values of the emissivity index $q$ are for the emissivity profile that falls off below the ISCO, but the actual value used is negative. The green lines and points mark the injected parameter values. \label{fig:Joa13ringratio}}
\end{figure*}

Studying the ratio of the subring radii may be of interest because it can possibly break some of the degeneracies present when analyzing a single subring. It is possible the ratio would be different for different values of spin and deformation from Kerr and so be a quite useful observable. The ratio of the subrings is mass independent since each subring's radius is linearly dependent on the mass. As with the subring radius observable, it is not possible to place any constraints at the lower resolution of $\sim 10~\mu$as. Figure~\ref{fig:Joa13ringratio} shows the posterior probabilities for the subring radius ratio observable for the ratio of the $n=0$ and $n=1$ subring radii and the $n=1$ and $n=2$ subring radii at the higher resolution of $\sim 1~\mu$as. Again the injection is Kerr and the model is the Johannsen spacetime with a non-zero $\alpha_{13}$ deformation parameter and the results are consistent across different deformation parameters and the two spacetimes. The ratio $n=0/n=1$ shows some promising results. The deformation parameter and spin are all better constrained than for the subring radius observable, though not as well as when analyzing the full shadow image. The emissivity index is constrained to a similar level and the inclination angle worse as compared with the subring radius observable. Of course the $n=0$ subring is strongly depend on the accretion disk and so these results may not hold with a more realistic disk, but these results do suggest looking at the ratio of the subring radii rather than just the subring radii could be useful for determing the properties of the black hole spacetime. The higher order ratio of $n=1/n=2$ does worse, however, and constraints cannot be placed on any parameters except the emissivity index. This is likely due to the difference in the radii of the $n=1$ and $n=2$ subrings being comparable to the resolution. A higher resolution would be required to make better use of this subring ratio.

%

\section{Conclusion}
\label{sec:conc}

This work has presented a study of the ability to place constraints on departures from the Kerr black hole solution and test general relativity using black hole shadow observations, with particular focus on the shadow subrings. Two different parametrized non-Kerr black hole metrics were used to model possible departures from Kerr and a simple accretion disk model with different emissivity profiles was used to model the disk. A large array of black hole shadow images was generated for a range of spins, deformation parameters, inclination angles, masses, and emissivity profiles. A basic likelihood analysis was performed to determine if and how well parameters could be estimated when the injection was a Kerr black hole and the model was some non-Kerr spacetime. This analysis was performed for two different resolutions, $\sim 10~\mu$as and $\sim 1~\mu$as, which are representative resolutions for earth-based and space-based versions of the EHT, respectively. The full image of the shadow, the locations of the subrings, and the ratio of the subring locations were each studied as distinct observables, both with and without the $n=0$ subring (i.e.~the direct image of the disk) included.

The results of the likelihood analysis suggest that shadow observations made with the lower resolution expected from earth-based VLBI will, at best, be able to place weak constraints on non-Kerr spacetimes. Additionally, due to degeneracies between some of the parameters, the parameter estimates may be biased and incorrect. The results are similar at the higher resolution when only the location of the subrings is studied. It is quite possible that the simple accretion disk model, particularly that the radiation was treated achromatically, and the simple analysis, which doesn't consider the mechanics of VLBI, used here are leading to such a pessimistic result, and a more complex analysis would show otherwise (see~e.g.~\cite{Broderick:2021ohx, Gelles:2021oyd} for related studies in Kerr). But this work does point out that it is not a foregone conclusion that earth-based VLBI will be able to make shadow observations at high enough accuracy to make claims about the strong-gravity properties of black holes. Much work is still required and one must be quite careful when making any claims. More optimistically, the analysis of the full shadow image at the higher resolution suggests future space-based VLBI will be able to place very good constraints on black hole spacetime properties, comparable or better than what is currently possible with other observations. Again, these results are affected by the simplicity of the analysis, but they show promise for what can be done with VLBI in the future.

There are clear improvements that can be done to the work presented here. A more realistic disk model could be implemented, such as that used in~\cite{Ozel:2021ayr, Younsi:2021dxe}. A full general relativistic magnetohydrodynamics simulation would lead to the most accurate images, however such simulations are very computationally costly and it would not be feasible to do so for the wide range of physical parameters studied here. One could also process the simulated images through a synthetic data generation pipeline and using more advanced image reconstruction techniques (see~e.g.~\cite{Tiede:2020iif, Gelles:2021oyd}). Doing so would be more representative of what is produced by EHT observations. These are certainly worthwhile studies as the EHT continues to make observations and plans are made for upgraded versions of the EHT. Finally, one could repeat the study performed here, but targeting other observables of the black hole shadow images. One such observable is simply the central dark region caused by the presence of an event horizon. The size of this region and how dark it is compared to the brighter emission from the disk can be a signature of non-Kerr black holes or other exotic compact objects (see~e.g.~\cite{Olivares:2018abq}). Another observable that is related to the photon subrings is the photon ring autocorrelations, which are correlations of intensity fluctuations on the photon ring. With more images of the black hole shadow spaced out in time it may be possible to use the correlations to estimate the mass and spin of the black hole without resolving the photon ring itself~\cite{Hadar:2020fda}. This has yet to be studied in the context of non-Kerr spacetimes and would certainly be useful and interesting, providing another alternative for testing the Kerr hypothesis and general relativity.

\ack

The author thanks Alejandro C\'ardenas-Avenda\~no for useful discussion. This work was supported by the Teach@T{\"u}bingen and Research@T{\"u}bingen Fellowships. Some calculations used the computer algebra system MAPLE, in combination with the GRTENSORIII package~\cite{grtensor}. The author acknowledges support by the High Performance and Cloud Computing Group at the Zentrum f{\"u}r Datenverarbeitung of the University of T{\"u}bingen, the state of Baden-W{\"u}rttemberg through bwHPC and the German Research Foundation (DFG) through grant no INST 37/935-1 FUGG.

\section*{References}
\bibliography{biblio}
\bibliographystyle{iopart-num}

\end{document}